\documentclass[12pt]{article} 
\usepackage{epsfig}
\begin{document}

\begin{center}
\noindent {\Large \bf
 
Effect of energy losses and interactions\\ 

during diffusive shock acceleration:\\

applications to SNR, AGN and UHE Cosmic Rays\\

}
~\\[2em]
R.J. Protheroe \\
Department of Physics, The University of Adelaide\\
Adelaide, Australia 5005\\ rprother@physics.adelaide.edu.au
\end{center}

\centerline{\bf \underline{Abstract}}

I discuss the shape of the high energy end of the spectrum of
particles arising from diffusive shock acceleration in the
presence of (i) additional diffusive escape from the accelerator,
(ii) continuous energy losses, (iii) energy changes arising from
interactions.  The form of the spectrum near cut-off is sensitive
to these processes as well as to the momentum-dependence of the
diffusion coefficients and the compression ratio, and so the spectrum
of any radiation emitted by the accelerated particles may reflect
the physical conditions of the acceleration region.  Results
presented in this paper have applications in interpreting the
spectral energy distributions of many types of astrophysical
object including supernova remnants (SNR), active galactic nuclei
(AGN) and acceleration sources of ultra high energy cosmic rays
(UHE CR).  Except for extremely nearby sources, spectral features
imprinted on the spectrum of UHE CR during the acceleration
process will be largely eroded during propagation, but the
spectrum of UHE neutrinos produced in interactions of UHE CR with
radiation, both during cosmic ray acceleration and subsequent
propagation through the cosmic microwave background radiation,
contains sufficient information to determine the cut-off momentum
of the UHE CR just after acceleration for reasonable assumptions.
Observation of these UHE neutrinos by the Pierre Auger
Observatory may help in identifying the sources of the highest
energy cosmic rays.\\



\newpage

\section{Introduction}

For particle acceleration by electric fields induced by the
motion of magnetic fields $B$ (including those at astrophysical
shocks), the maximum rate of momentum gain by relativistic
particles of charge $Ze$ can be written (in SI units)
\begin{equation}
\left. {dp \over dt} \right|_{\rm acc} = \xi(p) Ze c B
\label{eq:max_acc}
\end{equation}
where $\xi(p) < 1$ and depends on the details of the acceleration
mechanism.  For example, in diffusive shock acceleration (DSA)
$\xi(p)$ will depend on the shock velocity and diffusion
coefficients.  I shall refer to $\xi(p)$ as the acceleration rate
parameter.  Detailed and rigorous treatments of DSA are given in
several review articles
\cite{Drury83a,BlandfordEichler87,BerezhkoKrymsky88}; see
particularly the review by Jones \& Ellison \cite{JonesEllison91},
on the plasma physics of shock acceleration, which also includes a
brief historical review and refers to early work.

Protheroe and Stanev \cite{ProtheroeStanev98} proposed a box model
of DSA which was able to reproduce the essential features of DSA
in a simple leaky-box scenario.  Particles were injected into the
box at momentum $p_0$, and while inside the box their energy
increased at the {\em average} rate that would apply given the
physical parameters of the shock acceleration being modelled,
i.e.\ the upstream and downstream diffusion coefficients, the
shock velocity and compression ratio.  Particles being
accelerated leak out of the box, again at the {\em average} rate
that would apply given the physical parameters, and these
escaping particles represent the accelerated particles escaping
downstream in the standard shock acceleration picture.  This
simple scheme is very easy to implement in a numerical or Monte
Carlo program for investigating the structure of pile-ups and
cut-offs in the case of adding additional leakage terms (e.g.\
simulating the finite size of any accelerator), energy losses
during acceleration (e.g. synchrotron loss) or particle
interactions such as inverse Compton (IC) losses in the
Klein-Nishina regime.  These processes could be important in
determining the detailed shape of spectrum of non-thermal
radiation from SNR or AGN.

Where protons are accelerated, interactions with magnetic fields,
matter or radiation may determine where and how the spectrum cuts
off.  Of current interest is the hoped-for observation of $\pi^0$
TeV gamma rays from SNR from $pp$ interactions of cosmic rays
freshly accelerated at SNR shocks.  At higher energies, the UHE
CR may interact with background radiation fields, including the
cosmic microwave background radiation (CMBR), by pion
photoproduction interactions both during acceleration and
propagation.  This leads to a predicted spectral cut-off, the
Greisen-Zatsepin-Kuzmin (GZK) cut-off \cite{greisen,zatsepin} at
about $10^{20}$~eV for particles which have travelled more than a
few tens of Mpc from their sources.  However,
several experiments have reported UHE CR events with energies
above $10^{20}$~eV \cite{rf:TAK} with the highest energy event
having $3 \times 10^{20}$~eV \cite{rf:FE3}, although the latest HiRes
data \cite{HiRes02} seems to show a cut-off consistent with the expected
GZK cut-off.

Information about the acceleration environment is, in principle,
imprinted in the shape of the high energy end of the spectrum of
accelerated particles.  In the case of the UHE CR, one of the
main applications of the present work is to see to what extent
this information remains, either in the cosmic ray spectrum, or
in the spectrum of secondary particles such as neutrinos produced
during both acceleration and propagation
\cite{ProtheroeJohnson95,Stanevetal00,Mannheimetal01}.  Hence it
is important to calculate these fluxes resulting from cascading
in the CMBR.  Spectral data at energies above $10^{18}$eV and
directional results, notably from the AGASA project, are very
suggestive of fascinating, unexpected physics
\cite{Hayashidaetal99}.  A new era in the field is beginning with
the commissioning of the Pierre Auger Observatory
\cite{AugerCollaboration2001} which comprises a pair of
3000~km$^2$ arrays (one under construction in Mendoza Province,
Argentina, and one planned for Utah, U.S.A.) employing both
particle and optical detectors.  The Auger Observatory will also
have the capability of detecting UHE neutrinos which could carry
significant information about the sites of acceleration (see
refs. \cite{NaganoWatson00,ProtheroeClay04} for recent review of
UHE CR).

I shall first examine the case of continuous energy losses, then
consider how the spectral features are altered for the case of
interactions with various mean inelasticity values.  As a
specific example, I shall take the acceleration and propagation
of the UHE CR, and briefly re-examine the likely acceleration
sites of UHE CR.  That will be followed by a description of Monte
Carlo simulations of DSA using the box model and incorporating
Bethe-Heitler pair production and pion photoproduction
interactions in the CMBR; the results obtained being applicable also
to other temperature blackbody fields with suitable scaling.
Finally, I shall propagate these spectra to see whether any
useful features remain in the UHE CR spectrum, and whether or not the
spectrum of secondaries such as neutrinos can be used to infer
physical conditions in the acceleration region of UHE CR.

\section{Roll-offs and cut-offs in the spectrum of accelerated particles}

The inclusion of escape processes, in addition to the usual
escape downstream leads to a smooth roll-off in the spectrum of
accelerated particles which is most pronounced at momenta near the
momentum at which the two escape rates are equal, which I shall
refer to as the ``maximum momentum'', $p_{\rm max}$.  This smooth
roll-off occurs over up to three decades in energy and the
spectrum can extend beyond $p_{\rm max}$. Its shape depends on
the momentum dependence of the diffusion coefficients, on the
momentum dependence of the additional escape rate, and on the
intrinsic differential spectral index, $\Gamma$, of the
accelerated particles which would apply for the case of no energy
losses or additional escape processes.  Such roll-offs have very
recently been noted in test-particle Monte Carlo simulations of
shock acceleration at shocks in a cylindrical jet geometry where
there is sideways leakage out of the jet \cite{CasseMarkowith03}.

When continuous energy losses are included the spectrum is
cut-off sharply at a momentum at which the total rate of momentum
change is zero, and I shall refer to this as the ``cut-off
momentum'', $p_{\rm cut}$.  Depending on the intrinsic spectral
index, and momentum dependence of the diffusion coefficient, a
pile-up may occur before the cut-off.  Protheroe and Stanev \cite{ProtheroeStanev98}
subtracted from the momentum-gain rate a term representing
the energy-loss rate.  Drury et al.\ \cite{Druryetal99} showed that since the
physical size of the ``box'' increases with energy, synchrotron
losses can cause a particle in the downstream region to
effectively ``fall out'' of the box, and that this process could be
represented by an additional term in the formula for the escape
rate.  This term is included in the present work, and I confirm
the result of Drury et al.\ \cite{Druryetal99} that the pile-up is less,
particularly for low compression ratios.

In the box model of shock acceleration, particles of momentum
$p_0$ are injected into the acceleration zone, or ``box'', and
while inside the box are accelerated at a rate $r_{\rm acc}(p)$
and escape from the box at a rate $r_{\rm esc}(p)$.  These two
rates uniquely determine the spectrum of accelerated particles,
i.e.\ those escaping from the box.  The rate of change of
momentum during shock acceleration in the presence of energy
losses is
\begin{eqnarray}
pr_{\rm acc}(p) \equiv 
p\left[r_{\rm gain}(p)- r_{\rm loss}(p) \right]=\left.{dp \over dt}\right|_{\rm total}= \left. {dp
\over dt}\right|_{\rm gain}\! - \, \left. {dp \over dt}\right|_{\rm
loss}.
\end{eqnarray} 
Note that, by definition, $r_{\rm gain}(p_{\rm cut})=r_{\rm
loss}(p_{\rm cut})$.  

The momentum gain and escape rates depends on the diffusion
coefficients, which are usually assumed to have a power-law
dependence on momentum, $\kappa(p) \propto p^\delta$, with the
exponent depending on the nature of the turbulence present in the
magnetic field: $\delta=1/3$ (Kolmogorov spectrum), 1/2
(Kraichnan spectrum) or 1 (completely disordered field).  
When the diffusion coefficients upstream and downstream have the
same power-law dependence on momentum, $\kappa_1(p) = K_1
p^\delta$ and $\kappa_2(p) =K_2 p^\delta$, the momentum gain rate is
$r_{\rm gain}(p) \propto p^{-\delta}$, giving
\begin{eqnarray}
r_{\rm acc}(p) = r_{\rm loss}(p_{\rm cut})\left({p \over p_{\rm cut}}\right)^{-\delta} - r_{\rm loss}(p).
\end{eqnarray}
In the case of $dp/dt \propto p^2$ losses, such as synchrotron
loss, $r_{\rm loss}(p) \propto p$ and one has
\begin{eqnarray}
r_{\rm acc}(p) = r_{\rm loss}(p_{\rm cut}) \left[\left({p \over p_{\rm cut}}\right)^{-\delta} - {p \over p_{\rm cut}}\right].
\end{eqnarray}

If there are no energy losses and no additional escape processes
a power-law spectrum results, and the integral spectral index ,
($\Gamma$-1), equals the ratio of the escape rate to the
acceleration rate.  In this case, the escape rate representing
escape downstream is $(\Gamma-1)r_{\rm loss}(p_{\rm
cut})({p/p_{\rm cut}})^{-\delta}$.  The acceleration zone extends
distances $L_1(p)=\kappa_1(p)/u_1$ and $L_2(p)=\kappa_2(p)/u_2$
upstream and downstream from the shock, respectively.  Adding an
extra escape term $r_{\rm esc}^{\rm max}(p,p_{\rm max})$
responsible for the ``maximum momentum'', and an escape term
$r_{\rm esc}^{\rm fallout}(p)$ resulting from energy losses
causing particles to fall out of the box gives the total escape
rate
\begin{eqnarray}
r_{\rm esc}(p) =  (\Gamma-1)r_{\rm loss}(p_{\rm cut})\left({p \over p_{\rm cut}}\right)^{-\delta} + r_{\rm esc}^{\rm max}(p,p_{\rm max}) + r_{\rm esc}^{\rm fallout}(p) 
\end{eqnarray}
For the case of a constant extra escape rate as adopted by \cite{ProtheroeStanev98}
\begin{eqnarray}
r_{\rm esc}^{\rm max}(p,p_{\rm max}) =  (\Gamma-1)r_{\rm loss}(p_{\rm cut})\left({p_{\rm max}\over p_{\rm cut}}\right)^{-\delta},
\end{eqnarray}
and for the case of diffusive escape out of the box,
e.g.\ diffusion perpendicular to the shock normal to the edge of
the acceleration region with diffusion coefficient $\kappa
\propto p^\delta$,
\begin{eqnarray}
r_{\rm esc}^{\rm max}(p,p_{\rm max}) =  (\Gamma-1)r_{\rm loss}(p_{\rm cut})\left({p_{\rm max}^2 \over p \, p_{\rm cut}}\right)^{-\delta}.
\label{eq:sideways}
\end{eqnarray}
This latter case is probably more realistic than constant escape,
and will give weaker roll-offs.

The escape rate resulting from energy losses is
\begin{eqnarray}
r_{\rm esc}^{\rm fallout}(p)= {1 \over L(p)} {d L_2 \over dp}\left.{d p \over dt}\right|_{\rm loss} 
\end{eqnarray}
where $L(p)=L_1(p)+L_2(p)$, and only the decrease of $L_2(p)$
with decreasing momentum contributes to escape as particles
falling out of the upstream part of the box (length $L_1(p)$)
will be advected back into the box.  For $\kappa_1(p) = K_1
p^\delta$ and $\kappa_2(p) =K_2 p^\delta$,
\begin{eqnarray}
r_{\rm esc}^{\rm fallout}(p)= \delta \,\ell_2\, r_{\rm loss}(p)
\end{eqnarray}
where $\ell_2 \equiv {L_2(p) / L(p)}$ is related to the
compression ratio $R=u_1/u_2=(2+\Gamma)/(\Gamma-1)$,
\begin{eqnarray}
\ell_2 = {K_2 R \over K_1 + K_2 R} 
= {2 + \Gamma \over (K_1/K_2)(\Gamma -1)+2+\Gamma}.
\end{eqnarray}
Drury et al.\ \cite{Druryetal99} appear to use $K_2=K_1$
(possibly to represent a parallel shock for which in an ideal
case $B_2$=$B_1$), giving $\ell_2 = (2 + \Gamma)/(1+2\Gamma)=4/5$
for a strong shock. For a very weak shock $\ell_2 \to 1/2$.  For
a perpendicular shock, for which in an ideal case $B_2=RB_1$, one
might expect $K_2$=$K_1/R$, giving $\ell_2 = 1/2$.  Except where
otherwise stated, I shall use $\ell_2 = 1/2$ in this paper.

\section{Spectrum for cut-off due to continuous losses}

\begin{figure}[htb]
\centerline{\epsfig{file=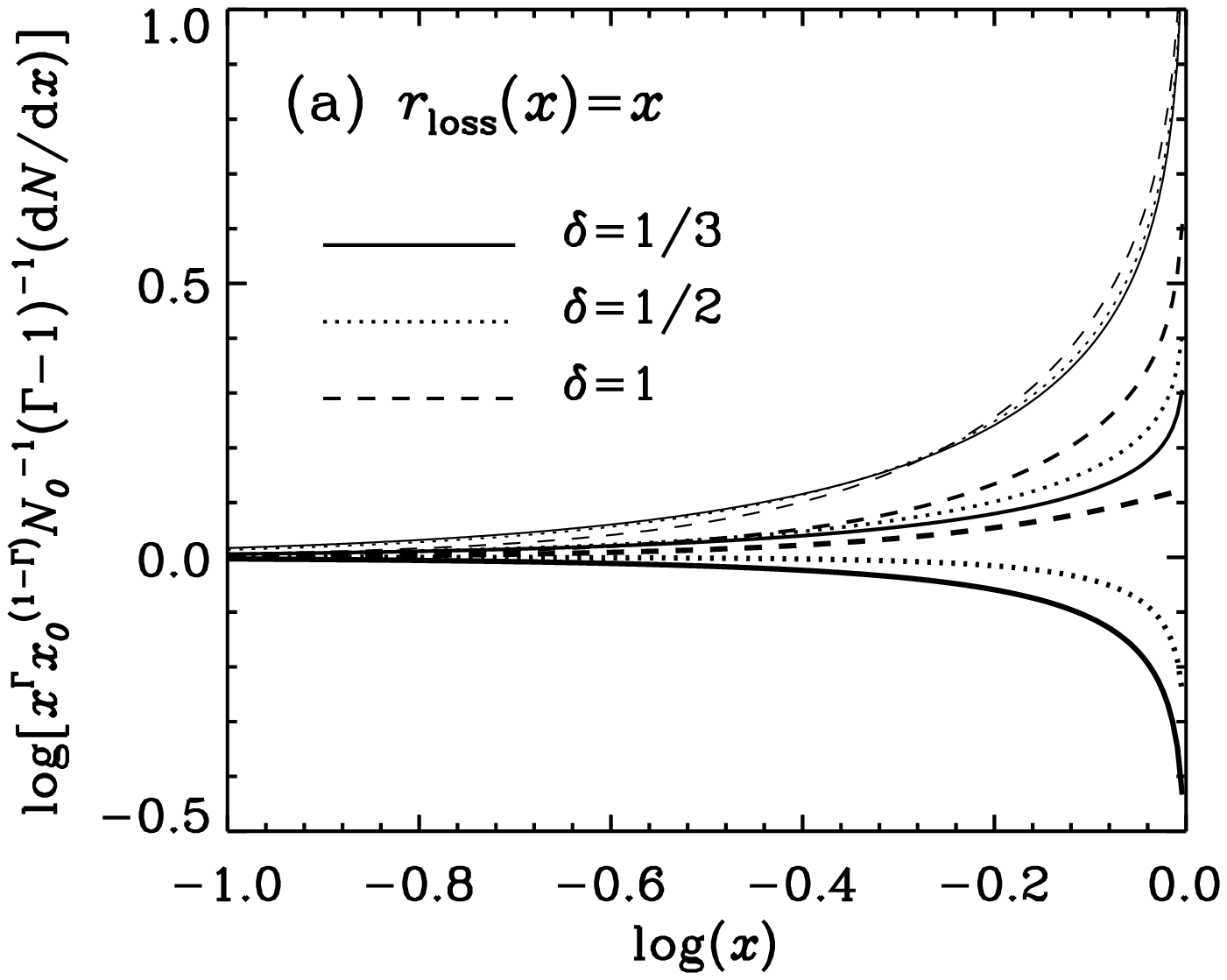,width=110mm}\hspace*{-4em}\epsfig{file=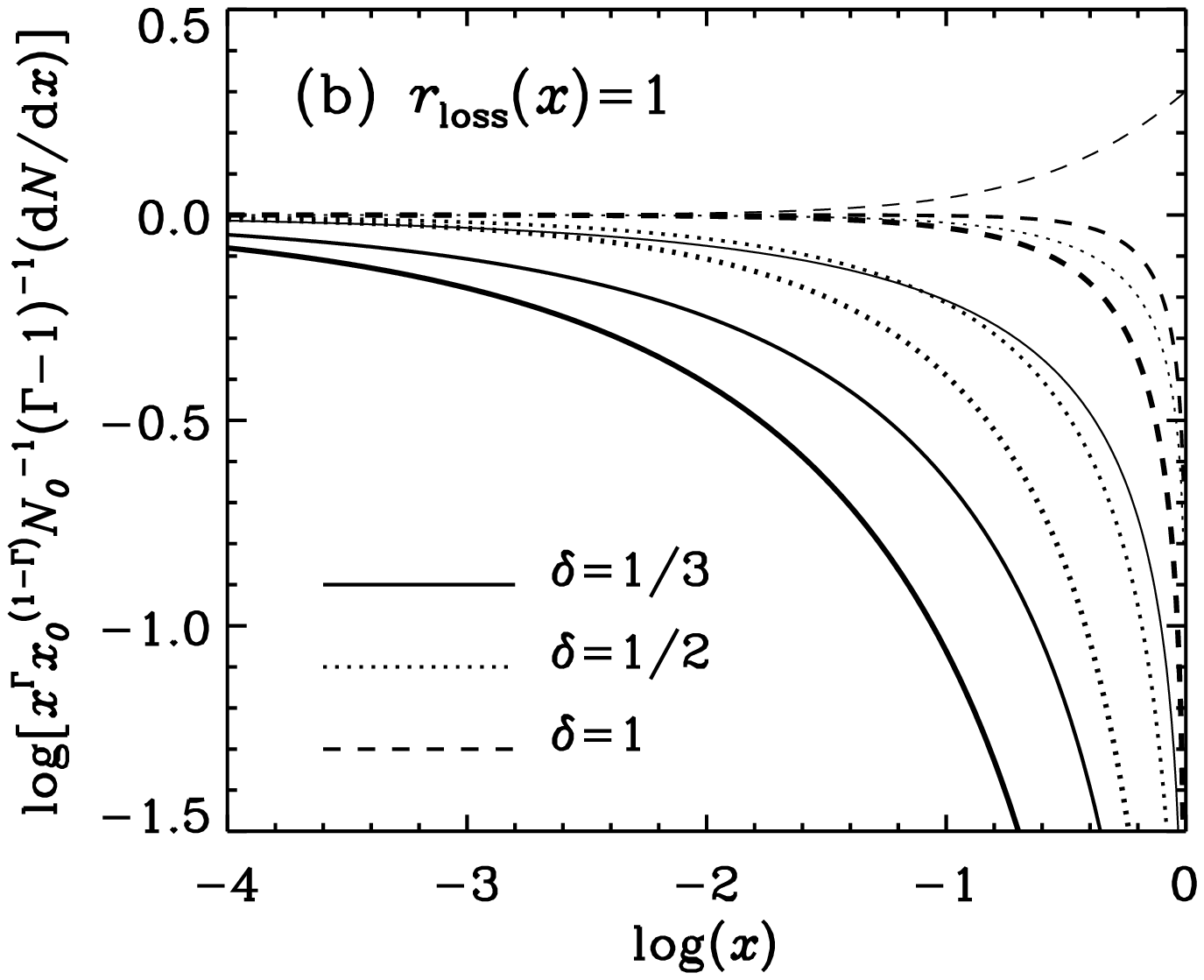,width=110mm}}
\caption{Differential energy spectra for $x_0 \ll 1 \ll x_{\rm
max}$ for $\ell_2=1/2$, $\delta$ as indicated, and $\Gamma=1.5$
(upper curves), 2.0 (middle curves) and 2.5 (lower
curves). Results are shown for (a) $dp/dt \propto p^2$, (b) $dp/dt
\propto p$.}
\label{synch_pileup}
\end{figure}

At time $t$ after injecting $N_0$ particles of momentum $p_0$ into the acceleration zone, the number
of particles remaining in the acceleration zone, $N(t)$, is obtained by solving
\begin{eqnarray}
{dN \over dt} = -N(t) r_{\rm esc}[p(t)]
\end{eqnarray}
which has solution
\begin{eqnarray}
N(t)=N_0 \exp \left[ - \int_0^t r_{\rm esc}[p(t)] dt \right] = 
N_0 \exp \left[ - \int_{p_0}^p r_{\rm esc}(p) {dt \over dp}dp\right] .
\end{eqnarray}
Hence, the spectrum of escaping particles is 
\begin{eqnarray}
F_{\rm esc}(p) &=& {dN_{\rm esc} \over dp} =  -{dN \over dt} {dt \over dp}
= N_0 {r_{\rm esc}(p) \over p \, r_{\rm acc}(p)}
\exp \left[ - \int_{p_0}^p { r_{\rm esc}(p)  \over p \, r_{\rm acc}(p)}dp\right] .
\label{eq:fund}
\end{eqnarray}
For continuous $dp/dt \propto p^2$ losses (synchrotron loss or IC
scattering in the Thomson regime), and defining $x\equiv p/p_{\rm cut}$,
from Eq.~\ref{eq:fund} the spectrum for the case of $x_0 \ll x <
1 \ll x_{\rm max}$ is given by
\begin{eqnarray}
F_{\rm esc}(x) \approx {\delta \ell_2 x + (\Gamma-1)x^{-\delta} \over x^{1-\delta} - x^2}
\left({x \over x_0}\right)^{-(\Gamma-1)}\left({1-x^{1+\delta}}\right)^{(\Gamma-1+\delta \ell_2)/(1+\delta)}.
\end{eqnarray}
For continuous $dp/dt \propto p$ losses (e.g., adiabatic), from
 Eq.~\ref{eq:fund} the spectrum for the case of $x_{\rm max} \gg
 1$ is given by
\begin{eqnarray}
F_{\rm esc}(x) \approx {\delta \ell_2 x + (\Gamma-1)x^{-\delta} \over x^{1-\delta} - x}
\left({x \over x_0}\right)^{\delta \ell_2} \left({x^{-\delta}-1 \over x_0^{-\delta}-1}\right)^{(\Gamma-1)/\delta + \ell_2}.
\end{eqnarray}
Note that in both cases, for $x \ll 1$ this reduces to $F_{\rm
esc}(p) \approx (\Gamma-1) x_0^{\Gamma-1} x^{- \Gamma}$, as it
must.  For $x_{\rm max} \to \infty$, and $\Gamma=1.5$, 2.0 and
2.5 the shape of the cut-off is shown in Fig.~\ref{synch_pileup}.
For synchrotron losses (Fig.~\ref{synch_pileup}a) the spectrum is
hardly affected up to $x=0.3$ and strong pile-ups are present for
$\Gamma \le 2$, but for the case of adiabatic losses
(Fig.~\ref{synch_pileup}b) a pile-up occurs only for $\Gamma <2$
and $\delta=1$.  Furthermore, adiabatic losses cause a much
stronger turnover and affects the spectrum at significantly
lower energies, down to $x \sim 10^{-3}$ for
$\delta=1/3$.  Of course the situation is modified for finite
$x_{\rm max}$.  In that case, the integral in Eq.~\ref{eq:fund}
involves Hypergeometric Functions (as used in
ref. \cite{ProtheroeStanev98}) and these are obtained by
summation of an infinite series.
  
\begin{figure}[htb]
\centerline{\epsfig{file=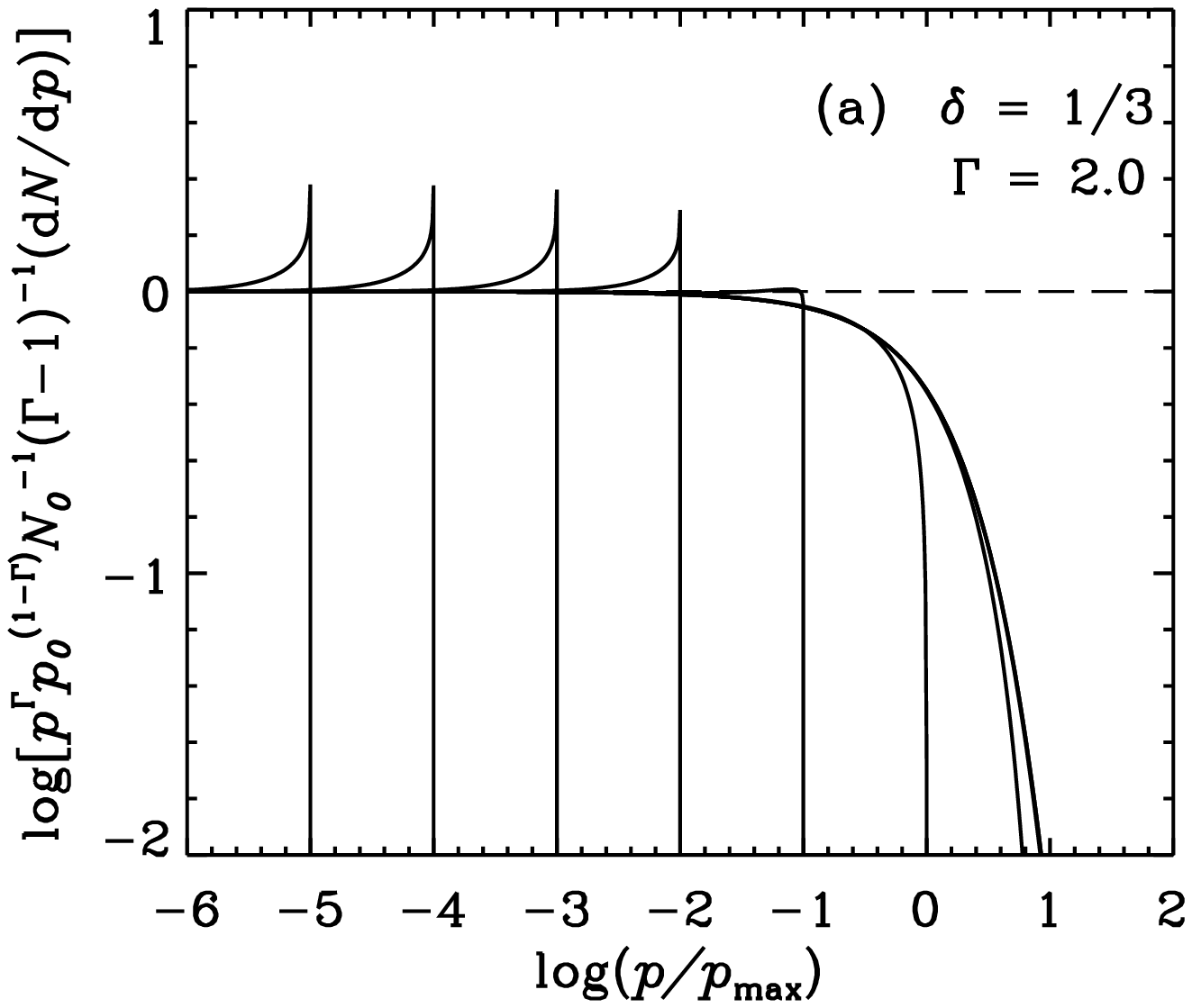,width=110mm}\hspace*{-4em}\epsfig{file=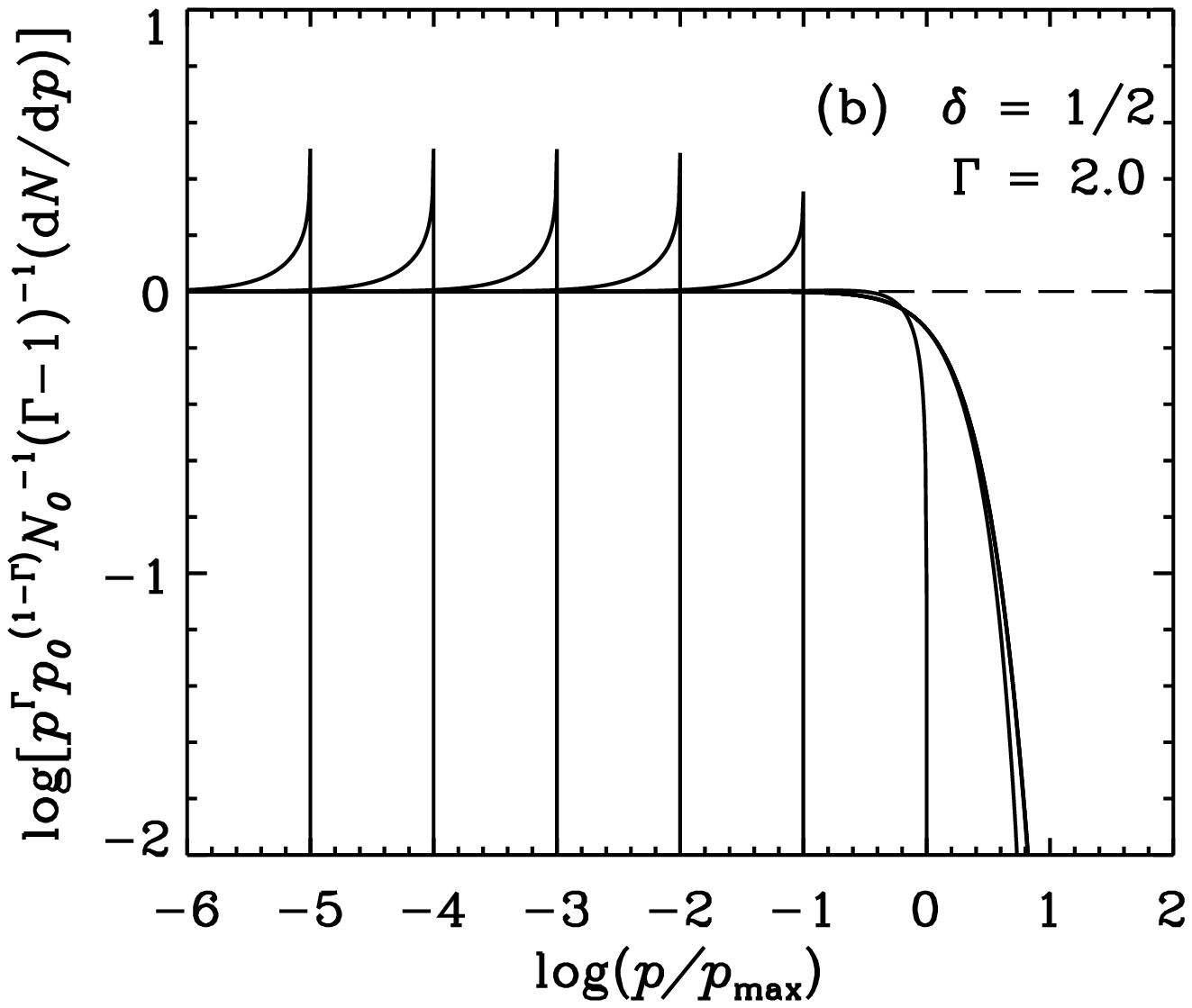,width=110mm}}
\centerline{\epsfig{file=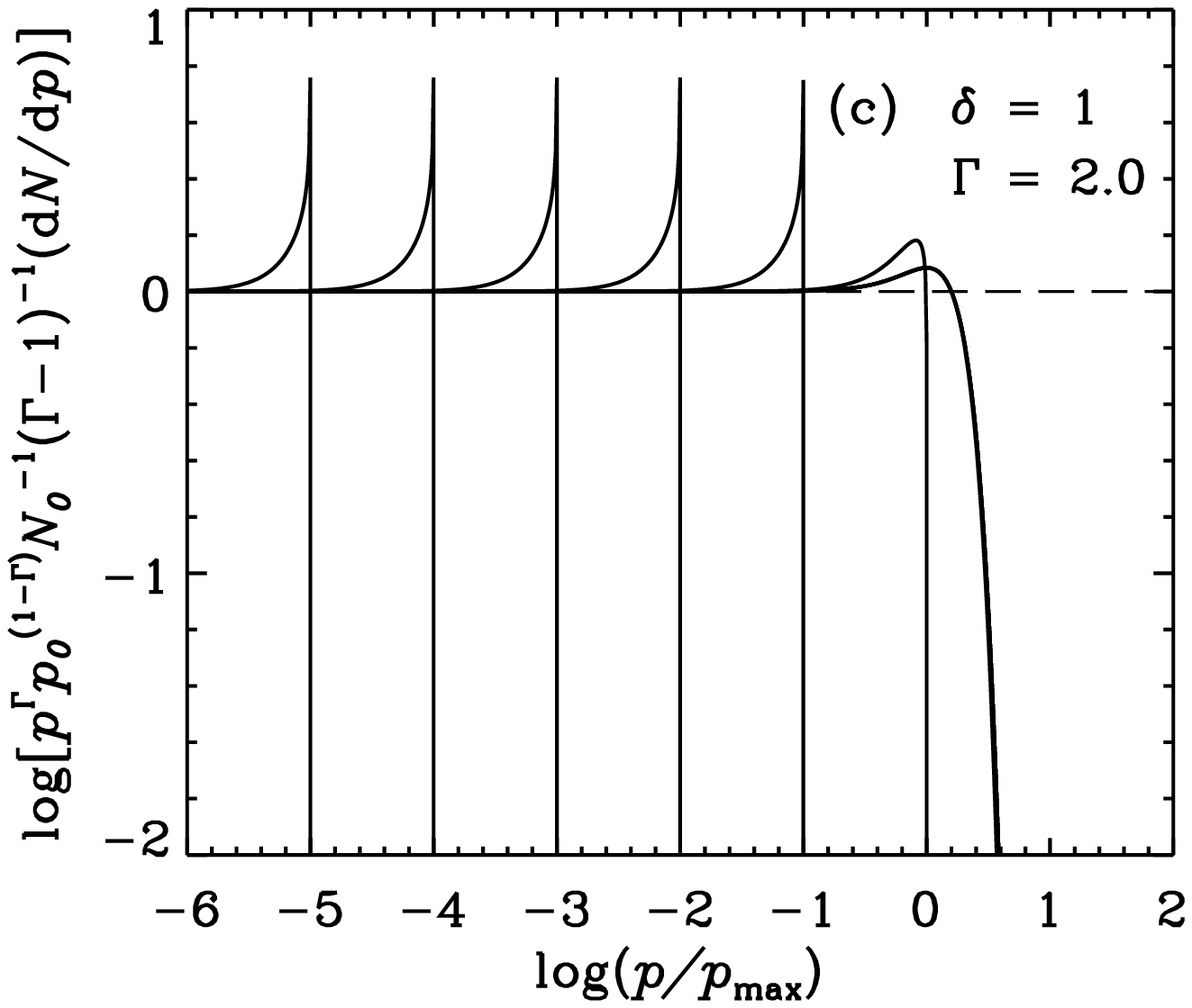,width=110mm}\hspace*{-4em}\epsfig{file=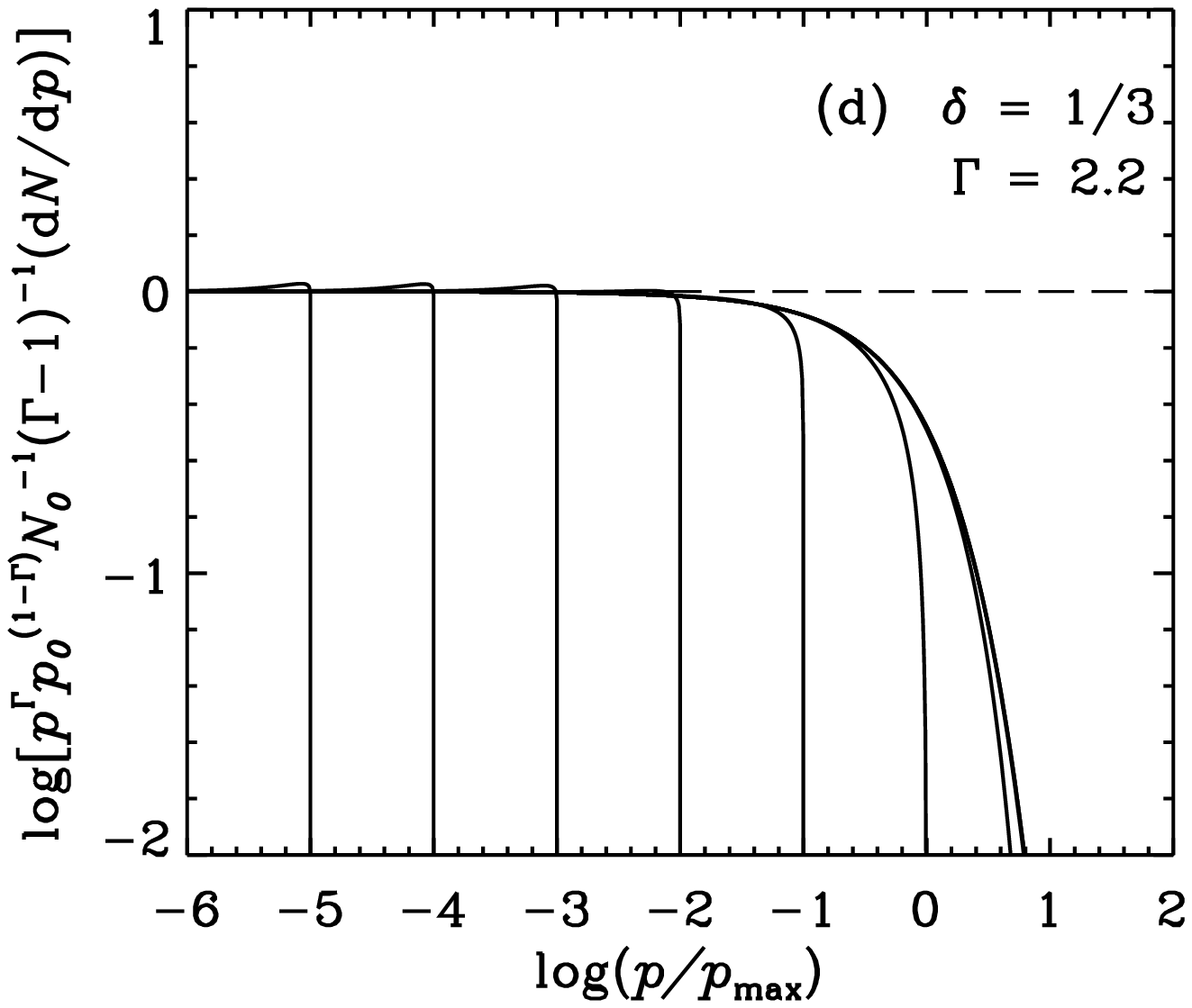,width=110mm}}
\caption{Differential energy energy spectrum for the case of
$\Gamma=2$ for (a) $\delta=1/3$, (b) $\delta=1/2$, (c)
$\delta=1$, and for (d) $\Gamma=2.2$ and $\delta=1/3$.  In each
case $\ell_2=1/2$, and results are shown for momentum-dependent escape at $p_{\rm
max}$ and $p_{\rm cut}/p_{\rm max} = 10^{-5}$ (leftmost curve),
$10^{-4}$, \dots , $10^{1}$ (rightmost curve).  }
\label{synch_cutoff_sideways}
\end{figure}

The spectral index as a function of momentum is given by
\begin{eqnarray}
{d \ln F_{\rm esc} \over d \ln p} &=& {p \over r_{\rm esc}(p)}{d \, r_{\rm esc} \over d \, p}
- {p \over r_{\rm acc}(p)}{d \, r_{\rm acc} \over d \, p} - 1 - {r_{\rm esc}(p) \over r_{\rm acc}(p)}.
\label{eq:fund_index}
\end{eqnarray}
One can obtain (the logarithm of) the differential energy spectrum by
integrating Eq.~\ref{eq:fund_index} numerically from $\ln p_0$ to
$\ln p$.  For the case of inclusion of both
synchrotron losses plus an extra momentum-dependent escape term
(Eq.~\ref{eq:sideways}), Eq.~\ref{eq:fund_index} gives
\begin{eqnarray}
{d \ln F_{\rm esc} \over d \ln x}  &=& {\delta \ell_2 x + \delta \, (\Gamma-1)(x^\delta  x_{\rm max}^{-2\delta}-x^{-\delta})  \over \delta \ell_2 x + (\Gamma-1) (x^{-\delta} \, +  \, x^\delta  x_{\rm max}^{-2\delta})} +{ \delta  x^{-\delta} + x  \over x^{-\delta} - x} \, \nonumber \\ &&    - \,  1 \, - \,  {\delta \ell_2 x + (\Gamma-1) (x^{-\delta} + x^\delta  x_{\rm max}^{-2\delta}) \over x^{-\delta} - x},
\label{eq:slope_syn_sideways}
\end{eqnarray}
and results for various $x_{\rm max}$, $\delta$ and $\Gamma$
values are shown in Fig.~\ref{synch_cutoff_sideways}.  These
results apply equally to electrons and protons if synchrotron
radiation is the dominant energy loss process.  An application in
which the cut-off in the proton energy spectrum is due to
synchrotron loss is given by the Synchrotron Proton Blazar model
for the jets of BL~Lac type active galactic nuclei which have
weak accretion disks \cite{MueckeProtheroe01,Mueckeetal03}.  In
contrast, in quasars with strong accretion disks, the spectrum of
accelerated protons may be cut off by pion photoproduction
interactions on thermal photons from the accretion disk.
Cut-offs due to pion photoproduction will be discussed in
Section~6.

Finally, one can find the criterion for a pile-up by examining
the spectral index as $x \to 1$ \cite{Druryetal99}.  The
denominators of the 2nd and 4th terms of
Eq.~\ref{eq:slope_syn_sideways} tend to zero as $x \to 1$, and so
the spectral index becomes infinite.  The sign of the denominator
is always positive if $\delta > -1$, and so whether the spectral
index tends to $+\infty$ or $-\infty$ depends on the sign of the
numerator.  Hence a strong pile-up occurs if
\begin{eqnarray}
\delta ( 1 -  \ell_2) +2 - \Gamma - (\Gamma-1)x_{\rm max}^{-2\delta} > 0.
\end{eqnarray}
Taking $x_{\rm max} \to \infty$ and
$\ell_2=(2+\Gamma)/(1+2\Gamma)$, as assumed by Drury et al.\ \cite{Druryetal99}, I
obtain
\begin{eqnarray}
(2 - \delta) +(3 +\delta)\Gamma -2\Gamma^2 > 0
\end{eqnarray}
from which the maximum values of $\Gamma$ for a pile-up are 2.069
($\delta=1/3$), 2.106 ($\delta=1/2$), and 2.225 ($\delta=1$).
Taking the compression ratio to be $R=(2+\Gamma)/(\Gamma-1)$, for
$\delta=1$ a pile-up is expected if $R>3.44949$ (in agreement
with Drury et al.\ \cite{Druryetal99}).  For $\ell_2$=1/2, as adopted in the present
paper, the conditions for pile-ups for synchrotron losses and
adiabatic losses are $\Gamma < 2+\delta/2$ and $\Gamma <
1+\delta$, respectively, and are seen to be consistent with the spectra
in Fig.~\ref{synch_pileup}.

\section{Spectrum for cut-off due to interactions}

Interactions of relativistic particles with matter or radiation
generally result in the interacting particle losing energy,
which is either given to the struck particle or photon (elastic
collision) or used in production of secondary particles
(inelastic collision).  If the mean interaction length (mean free
path) is $\lambda_{\rm int}(p)$, for ultra-relativistic particles
the rate of interaction is $r_{\rm int}(p)=c/\lambda_{\rm
int}(p)$.  If the mean inelasticity, i.e.\ average fraction of
energy lost per interaction, is $\bar{\alpha}(p)$ then the
effective loss rate for these non-continuous losses is $r_{\rm
loss}(p)=\bar{\alpha}(p)r_{\rm int}(p)$.  Of course, if
$\bar{\alpha}(p) \ll 1$ then one can approximate the interactions
as continuous energy losses.  I shall now investigate how the
spectrum of particles subject to interactions during acceleration
differs from that of particles subject to continuous
energy losses with the same effective loss rate.

In the simulation with $N_0$ particles injected, a particle is
injected at time $t$=0 with momentum $p(0)=p_0=x_0 p_{\rm cut}$
and statistical weight $w(0)=w_0=1/N_0$, and its subsequent
momentum, $p(t)=x(t) p_{\rm cut}$, and weight, $w(t)$, are
determined after successive time steps $\Delta t$ chosen to be
much smaller than the smallest time-scale in the problem; $\Delta
t = 0.01 x_0^{-\delta}/r_{\rm loss}(p_{\rm cut})$ is used in
these simulations.  In each time step, first the momentum is
changed, $p(t+\Delta t)=p(t)[1+\Delta t \, r_{\rm acc}(p)]$, and
then the probability of escaping in time $\Delta t$ is estimated
as $P_{\rm esc}=\{1-\exp[-\Delta t \, r_{\rm esc}(p)]\}$ where
$r_{\rm esc}$ here does {\em not} include the $r_{\rm esc}^{\rm
fallout}$ term. Then $w(t)P_{\rm esc}$ particles with momenta $p$
are binned in a histogram of accelerated particles, and the
particle's weight is changed to reflect the fraction not
escaping, $w(t+\Delta t)=w(t)(1-P_{\rm esc})$.  Next, the
probability of interacting in time $\Delta t$ is estimated,
$P_{\rm int}=\{1-\exp[-\Delta t \, r_{\rm int}(p)]\}$, and a
random number is generated to determine whether or not an
interaction takes place.  If an interaction does take place, a
random number is generated and used to sample the inelasticity
$\alpha$, and the particle's momentum ($p_i$ before the
interaction) is changed to $p = p_i(1-\alpha)$.  Now the
additional escape resulting from the momentum decreasing during
the interaction is considered. The probability of the particle
immediately escaping downstream due to ``falling out of the box''
is estimated,
\begin{eqnarray}
{\rm Prob.(escape},p_i \to p) = {L_2(p_i)-L_2(p) \over
L(p_i)} = \left[1-\left({p \over
p_i}\right)^\delta\right]\ell_2,
\end{eqnarray}
and a random number is generated to determine if this happens.
If the particle does escape, then $w(t+\Delta t)$ particles with
momenta $p$ are binned in a histogram of accelerated particles,
and a new particle is injected with momentum $p_0$ and weight
$w_0$.  If the particle does not escape, then the particle's
momentum and weight are evolved through a new time step as
described above.

\begin{figure}[htb]
\centerline{\epsfig{file=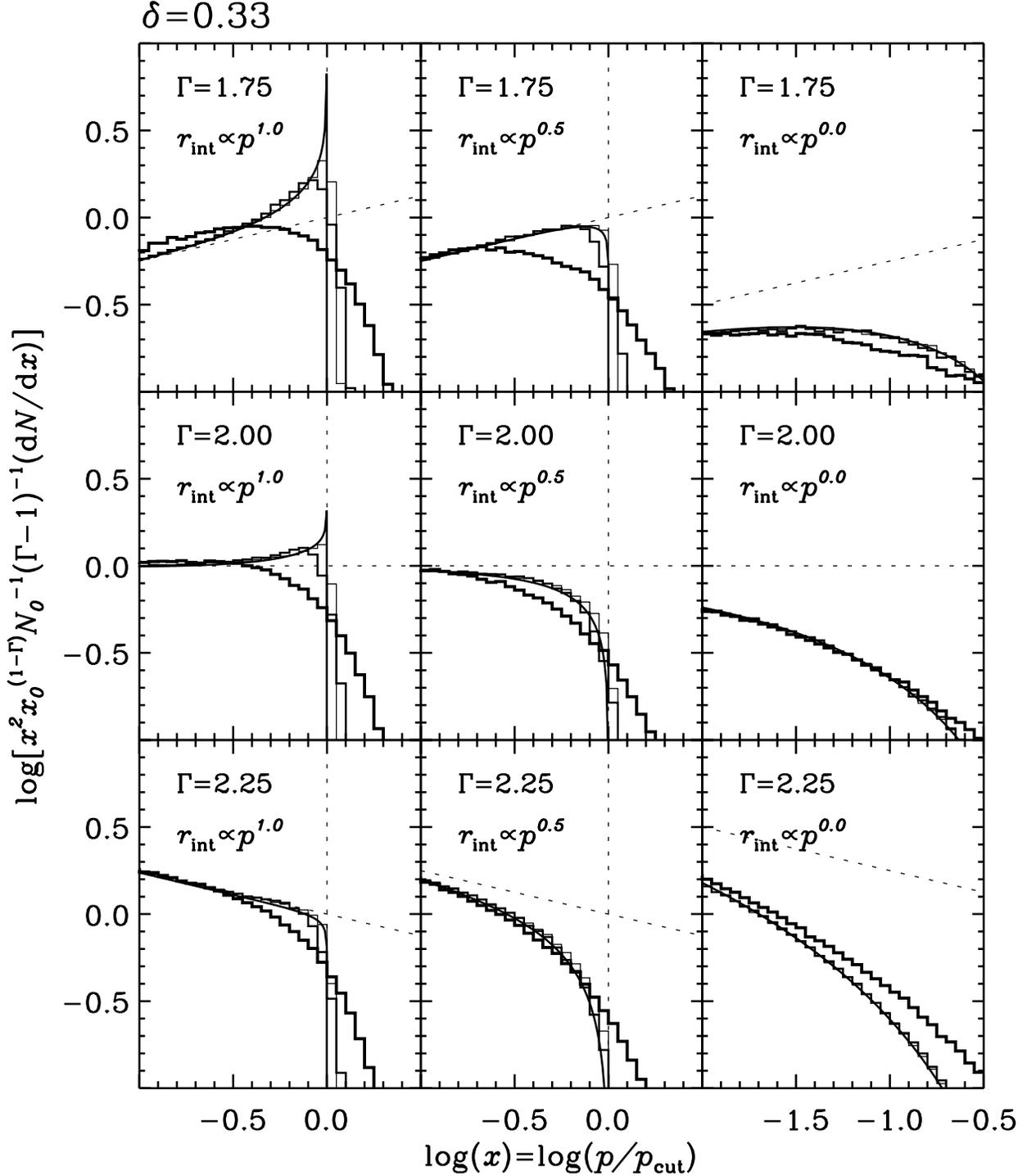,width=300mm}\hspace*{2cm}}
\caption{Spectra for $\delta=1/3$, $\ell_2=0.5$ and combinations
of $\Gamma$=1.75, 2 and 2.25, and $r_{\rm loss}(p)\propto
p^0,p^{0.5}$ and $p^1$ as indicated.  In each case the histograms show Monte
Carlo results for $\bar{\alpha}$=0.005 (thin histogram),
$\bar{\alpha}$=0.05 (intermediate histogram) and
$\bar{\alpha}$=0.5 (thick histogram), and the solid curve gives
the analytic result for continuous losses.}
\label{crude_a}
\end{figure}

\begin{figure}[htb]
\centerline{\epsfig{file=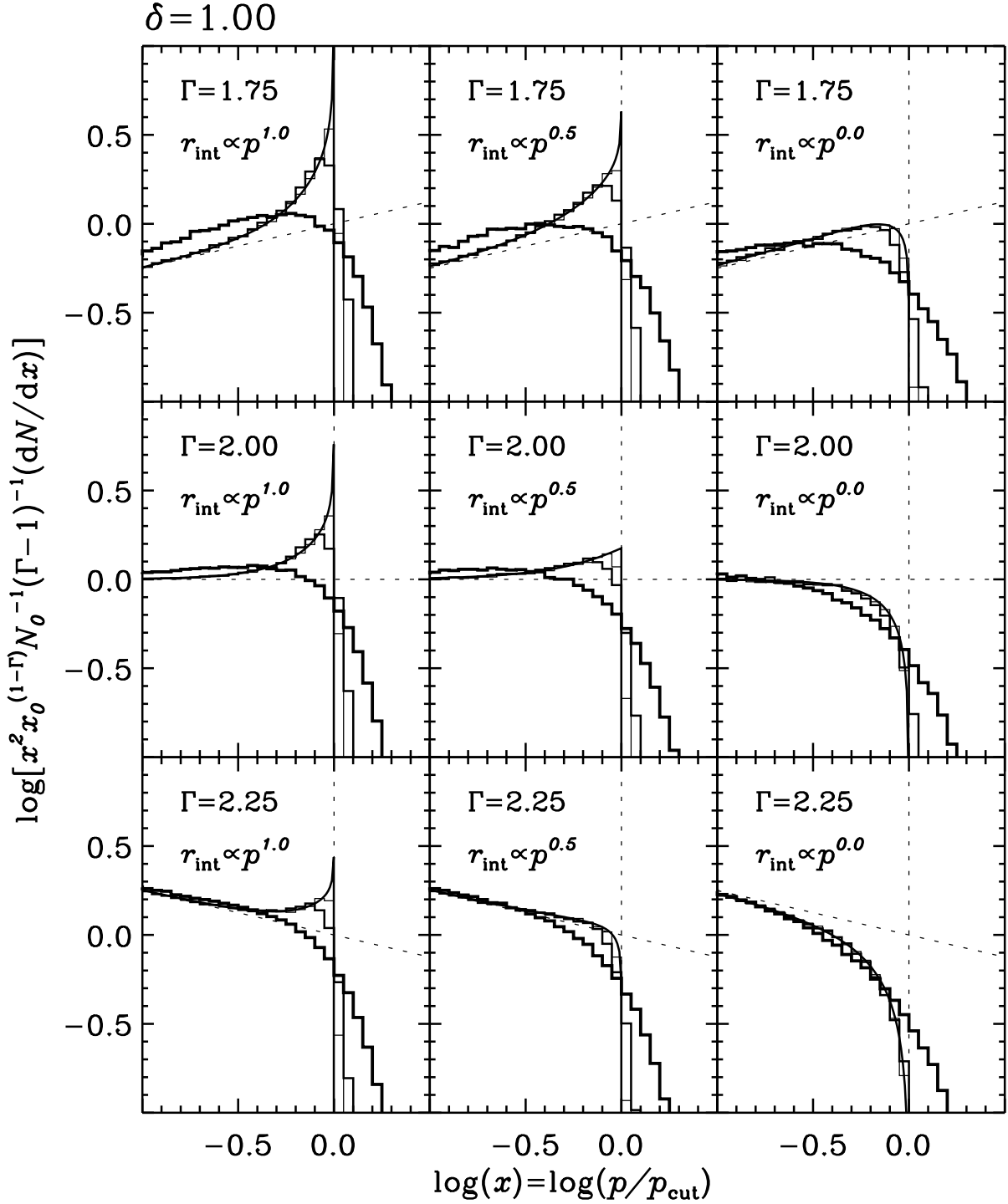,width=300mm}\hspace*{2cm}}
\caption{As Fig.~\protect\ref{crude_a} except $\delta=1$.}
\label{crude_c}
\end{figure}

The resulting spectra are given in Figs.~\ref{crude_a} and \ref{crude_c} give results
for $\delta$=1/3 and 1 for the cases for which we
already have results for the corresponding continuous loss
process, i.e. the same $r_{\rm loss}(p)$ but with $\bar{\alpha}
\to 0$, for $\bar{\alpha}=0.005,0.05$ and 0.5, and (for
simplicity) an inelasticity distribution which is constant for $0
< \alpha \le 2\bar{\alpha}$ and zero elsewhere.  In each figure I show spectra for
combinations of $\Gamma$=1.75, 2 and 2.25, and $r_{\rm
loss}(p)\propto p^0,p^{0.5}$ and $p^1$.  The histograms show
Monte Carlo results for the three inelasticity values, and the solid
curve gives the analytic result for continuous losses which is
found to be in excellent agreement with the $\bar{\alpha} =0.005$
case.  As is to be expected, where a strong pile-up (spike) is
present for the case of continuous losses, it remains but is
reduced if $\bar{\alpha}=0.05$, and is completely removed if
$\bar{\alpha}=0.5$.  In the last case, the spectrum generally
continues to momenta above $p_{\rm cut}$, and there can also be an
enhancement (weak pile-up) at lower momenta.

The spectra obtained above for the $\bar{\alpha}=0.5$ and $r_{\rm
loss}(p)\propto p^0$ would be directly applicable to the case
where proton acceleration is cut off as a result of hadronic
interactions with matter above the threshold for multiple
pion-production because the proton-nucleon inelastic cross
section increases only logarithmically with energy, and may be
approximated as constant over, say, one decade in energy.
Furthermore, for this case a not-unreasonable approximation for
the inelasticity distribution is a uniform distribution between 0
and 1.  Although other processes are normally assumed to cut off
the spectrum, if the matter density is sufficiently high, it is
in principle possible for hadronic collisions of protons with
matter to play this role.  Another possible application, again
for the $r_{\rm loss}(p)\propto p^0$ case, would be for
acceleration of electrons in relatively dense astrophysical
matter in which relativistic bremsstrahlung could cut off the
electron spectrum.  Although a flat inelasticity distribution is
not such a good approximation in this case, this process does
involve discontinuous energy losses comparable to the electron's
energy, and so the results for $\bar{\alpha}=0.5$ may serve as a
guide.

Perhaps, the most interesting application, again for the
$\bar{\alpha}=0.5$ and $r_{\rm loss}(p)\propto p^0$ case would be
for acceleration of high energy protons in a dense
radiation field to energies well above the pion photoproduction
threshold where $r_{\rm loss}(p)$ is approximately constant and
$\bar{\alpha}$ ranges from 0.2 to 0.5 depending on proton energy.
Near threshold $r_{\rm loss}(p)$ is strongly dependent on
momentum, and just above threshold the momentum dependence
changes smoothly and passes through $r_{\rm loss}(p)\propto p$
and $r_{\rm loss}(p)\propto p^{1/2}$ before becoming approximately
constant well above threshold.  For proton acceleration in a
radiation field where pion photoproduction cuts off the spectrum,
one can expect to see features of the type shown in
Figs.~\ref{crude_a}--\ref{crude_c} depending on the cut-off
momentum, and momentum dependence of the diffusion coefficient.
This process will be discussed in greater detail in the next
two sections for the case of black-body radiation.

\section{Cut-off momentum for acceleration in black body radiation: sites of UHE CR}

Although I shall concentrate here on cut-offs due hadronic
interactions with photons of the CMBR during acceleration of 
UHE CR protons, the results presented will be equally
applicable to proton acceleration in any black body radiation
field provided the proton momenta and interaction time-scales are
scaled appropriately.  For example, they may be applicable to
proton acceleration in quasars in the nearly black-body accretion
disk radiation.  For an arbitrary temperature $T=2.725\,
T_{2.7}\,$K, the momenta quoted should be replaced by
$p/T_{2.7}$.

\begin{figure}[htb]
\centerline{\epsfig{file=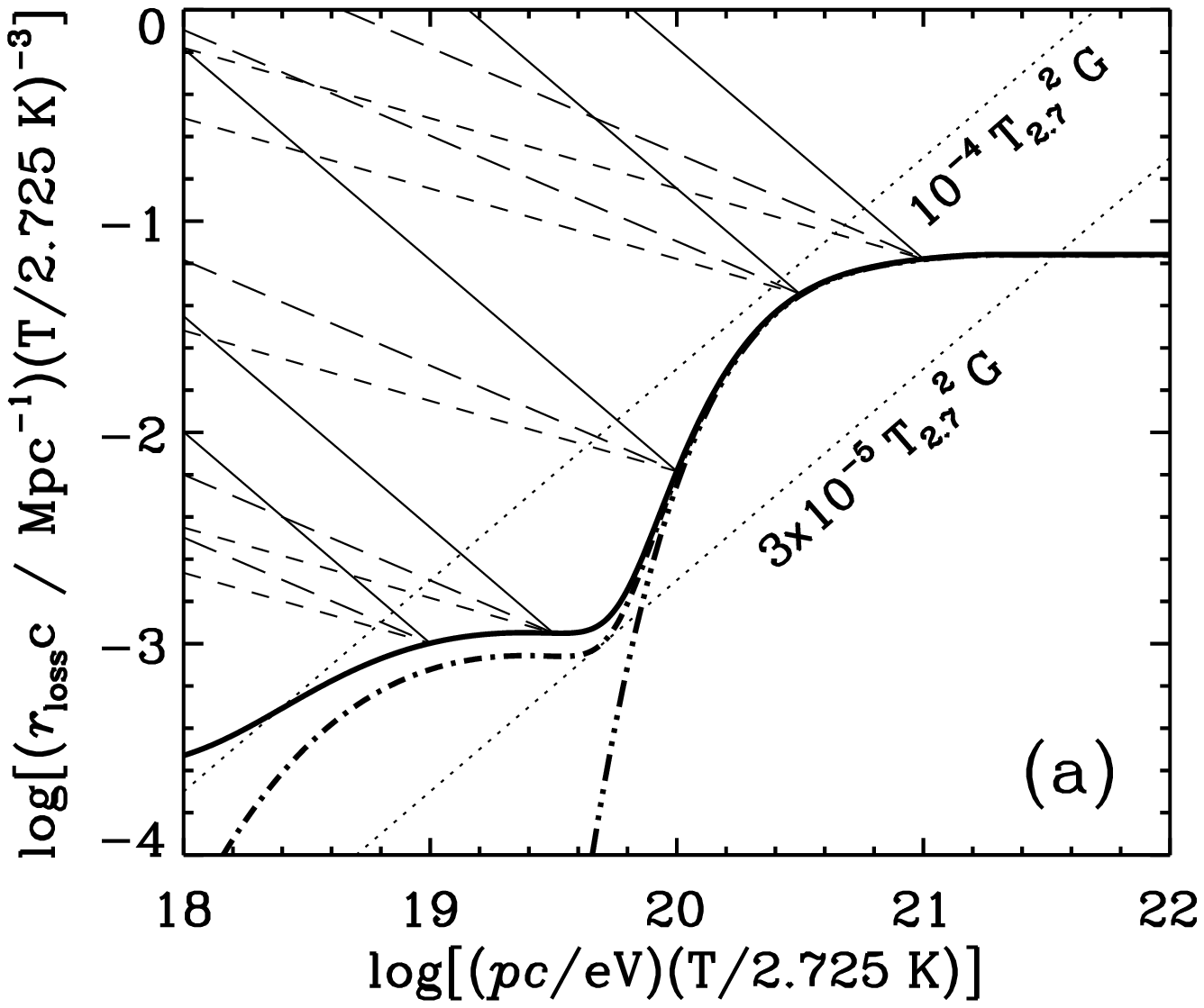,width=110mm}\hspace*{-4em}\epsfig{file=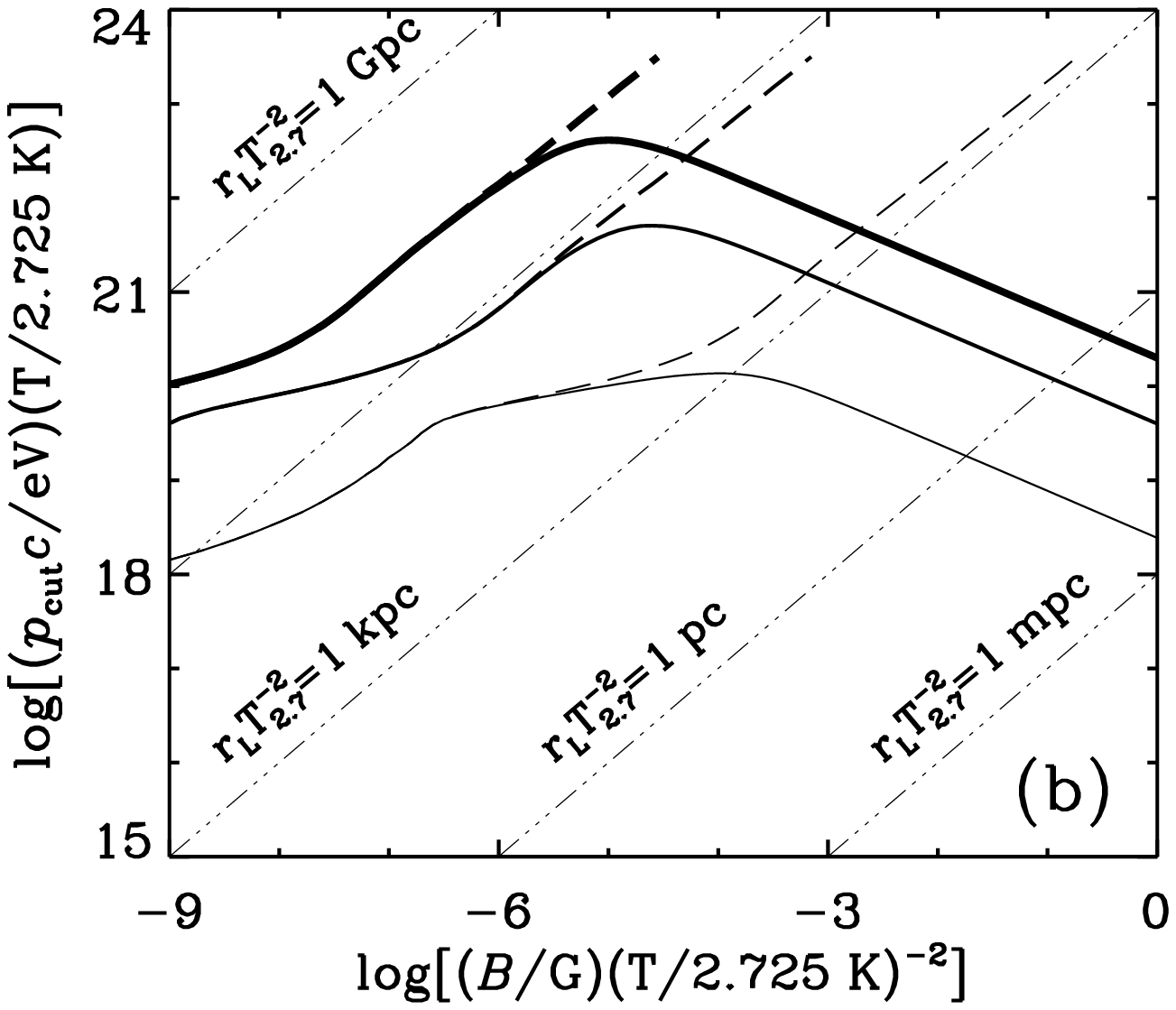,width=110mm}}
\caption{(a) Energy loss rate of protons $r_{\rm loss}(p)$ in
black body radiation of temperature $T=2.725\, T_{2.7}\,$K with
losses by Bethe-Heitler pair production, pion photoproduction and
redshifting \protect\cite{Stanevetal00}.  Note that redshift
losses are included to illustrate their importance in the case of
UHE CR acceleration in the 2.725~K CMBR.  For other temperatures,
the rate for Bethe-Heitler pair production plus pion
photoproduction (chain curve) should be used; the
dot-dot-dot-dash curve gives the loss rate for pion
photoproduction only.  Acceleration rates needed to reach cut-off
momenta $p_{\rm cut}c/T_{2.7}= 10^{19}$, $10^{19.5}$, \dots ,
$10^{21}$~eV are shown by the thin lines: short dashed
($\delta=1/3$), long dashed ($\delta=1/2$) and solid
($\delta=1$).  Dotted lines show synchrotron loss rates for
$B/T_{2.7}^2=3\times 10^{-5}$G and $10^{-4}$G.  (b) Maximum
energy as a function of magnetic field of protons for maximum
possible acceleration rate $\xi=1$ (upper solid curve), plausible
acceleration at perpendicular shock $\xi=0.04$ (middle solid
curve), and plausible acceleration at parallel shock $\xi=1.5
\times 10^{-4}$ (lower solid curve).  Dashed curves are limits
from Bethe-Heitler pair production and pion photoproduction only
(solid curves include synchrotron loss).  Dot-dot-dot-dash curves
are lines of constant Larmor radius as labelled.}
\label{photoprod_rate}
\end{figure}

In the case of acceleration of UHE CR, interactions with the CMBR
may be important if the magnetic field is less than $\sim
10^{-4}$~G.  In this case, the gyroradii are rather large, and
possible acceleration sites could be hot-spots in lobes of giant
radio galaxies \cite{Hillas84,RachenBiermann93}, or MHD generated
electric fields along the whole length of an AGN jet such as that
of Cen~A \cite{Schopperetal02}.  Inevitably, acceleration times will be long, and one
should also include redshift losses, and Bethe-Heitler pair
production which has a somewhat lower threshold than pion
photoproduction.  In Bethe-Heitler pair production, the energy
lost per interaction is small, less than 1\%, and so this process
can be treated adequately as a continuous loss process.  The
total momentum loss rate for these three processes is given in
Fig.~\ref{photoprod_rate}(a) together with the momentum-gain
rates required to reach various cut-off momenta for $\delta$=1/3,
1/2 and 1.

By plotting magnetic field vs.\ size of various astrophysical
objects, Hillas \cite{Hillas84} identified possible sites of acceleration
of UHE CR based on whether or not the putative source could
contain the gyroradius of the accelerated particles, and on the
likely velocity of scattering centres in these sites (arguments
similar to those that led to Eq.~\ref{eq:max_acc}).  Following
Hillas \cite{Hillas84} one finds that possible sites included neutron
stars ($10^7$--$10^{13}$G), gamma ray bursts and active galactic
nuclei ($10^3$--$10^{4}$G), and lobes of giant radio galaxies and
galaxy clusters ($10^{-7}$--$10^{-5}$G).  This identification of
possible sources does not take account of energy losses
(synchrotron) and interactions (Bethe-Heitler and pion
photoproduction) which can cut off the spectrum, and so
apply an additional constraint.  In the same paper, Hillas \cite{Hillas84}
used this additional constraint to narrow the field of possible sources to
radio galaxy lobes and galaxy clusters, and showed that to
accelerate protons to $\sim 10^{20}$eV large regions ($\sim$Mpc)
of relatively low magnetic field (less than $\sim 10^{-4}$G) are needed,
ruling out high magnetic field regions for the origin of UHE CR.
For the higher magnetic fields in this range the spectrum is
cut-off by synchrotron loss, and would be as discussed in
Section~3.  Thus, one of the very few plausible acceleration sites of
UHE CR may be associated with the radio lobes of powerful radio
galaxies, either in the hot spots \cite{RachenBiermann93} or
possibly the cocoon or jet \cite{Normanetal95}.  

To estimate cut-off momenta (or energy), one needs plausible
values for the acceleration rate.  The following
values~\cite{Protheroe00} for the acceleration rate parameter in
Eq.~\ref{eq:max_acc} are used: maximum possible acceleration rate
$\xi(p_{\rm cut})$=1, plausible acceleration at perpendicular
shock with speed 0.1$c$, $\xi(p_{\rm cut}) \approx 0.04$, and
plausible acceleration at parallel shock with speed 0.1$c$,
$\xi(p_{\rm cut}) \approx 1.5\times$$10^{-4}$.  The latter value
is consistent with that of Biermann and Strittmatter
\cite{BiermannStrittmatter87} who considered DSA of protons
in radiation and magnetic fields of active galactic nuclei.
Assuming a Kolmogorov spectrum of turbulence, equation~6 of
Biermann and Strittmatter \cite{BiermannStrittmatter87} would
give an acceleration rate parameter at the maximum energy of $\xi
= 0.08 \beta_s^2 b$ where $\beta_sc$ is the shock velocity and
$b$ is the ratio of turbulent to ambient magnetic energy density;
for $\beta_s$=0.1 this gives $\xi(p_{\rm max}) \approx
8\times$$10^{-4}b$.  Based on the total momentum loss rate given
in Fig.~\ref{photoprod_rate}(a), the proton cut-off momentum is
plotted in Fig.~\ref{photoprod_rate}(b) as a function of magnetic
field of protons for three adopted $\xi$-values (chain lines are for
constant Larmor radius as labelled).  This plot, which is based
on that in ref.~\cite{Protheroe00}, clearly shows that to
accelerate protons to $\sim 10^{20}$eV large regions of
relatively low magnetic field $\sim 10^{-7}$--$10^{-3}$G are
needed, apparently ruling out high magnetic field regions for the
origin of UHE CR (see also ref.~\cite{Medvedev03}).  One sees that, in
principle, protons can be accelerated up to $\sim 5 \times
10^{22}$eV in Mpc scale region with $\sim 10^{-5}$G.  

\begin{figure}[htb]
\centerline{\epsfig{file=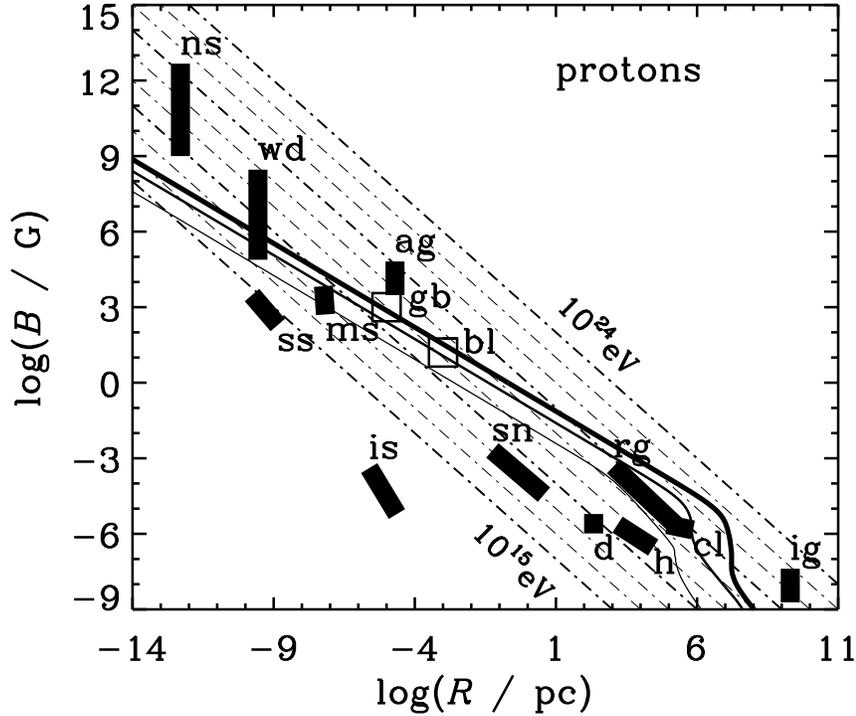,width=150mm}}
\caption{``Hillas plot'' showing (chain curves) magnetic field
vs.\ gyroradius for proton momenta $10^{15}$, $10^{16}$, \dots,
$10^{24}$eV/c.  The solid curves correspond to those in part
Fig.~\protect\ref{photoprod_rate}(b) such that the parameter
space of accelerated particles is to the left of the curve
corresponding to the chosen acceleration rate parameter. Typical
size and magnetic field of possible acceleration sites (taken
from Hillas \protect\cite{Hillas84}) are shown for neutron stars
(ns), white dwarfs (wd), sunspots (ss), magnetic stars (ms), AGN
(ag), interstellar space (is), supernova remnants (sn), radio
galaxy lobes (rg), galactic disk (d) and halo (h), clusters of
galaxies (cl) and intergalactic medium (ig).  Typical jet-frame
parameters of the synchrotron proton blazar model
\protect\cite{Mueckeetal03} and gamma ray burst model
\protect\cite{PelletierKersate00} are indicated by the open
squares labelled ``bl'' and and ``gb''.  }
\label{proton_emax}
\end{figure}

Fig.~\ref{proton_emax} is the ``Hillas plot'' with constraints
added corresponding to the three curves in
Fig.~\ref{photoprod_rate}(b), and where the chain lines give
constant proton energy values as indicated; sources to the right
of the solid curves being excluded.  A possible exception to
this is in the case of relativistically beamed sources
(e.g.\ for AGN see ref.\ \cite{Protheroeetal03}, and for GRB see
ref.\ \cite{PelletierKersate00}) where neutrons emitted along
the direction of relativistic motion can be Doppler boosted
significantly in energy.  Another possible exception is the case
of so called ``one-shot'' mechanisms (e.g.\
\cite{Haswelletal92,Sorrell87}) where a particle is accelerated by
an electric field along a nearly straight path which is
essentially parallel to the magnetic field such that curvature
and synchrotron losses are negligible.  Suggested sites include
polarization electric fields arising in plasmoids injected into a
neutron star's magnetosphere \cite{LitwinRosner01} and magnetic
reconnection in the magnetosphere of accretion induced collapse
pulsars \cite{deGouvelaDalPinoLazarian01}.  Another possibility
is plasma wakefield acceleration, i.e.\ acceleration by
collective plasma waves, possibly in the atmosphere of a GRB, or
``surf-riding'' in the approximately force-free fields of the
relativistic wind of a newly born magnetar \cite{Arons03}.  In
these cases it is unclear whether the requirements of negligible
radiation losses can be met.  I will next discuss in some detail
shape of the spectrum, where the cut-off is due to photoproduction
losses, for a large range of cut-off momenta.

\section{Spectrum for cut-off due to photoproduction loss}

\begin{figure}[htb]
\centerline{\hspace*{2cm}\epsfig{file=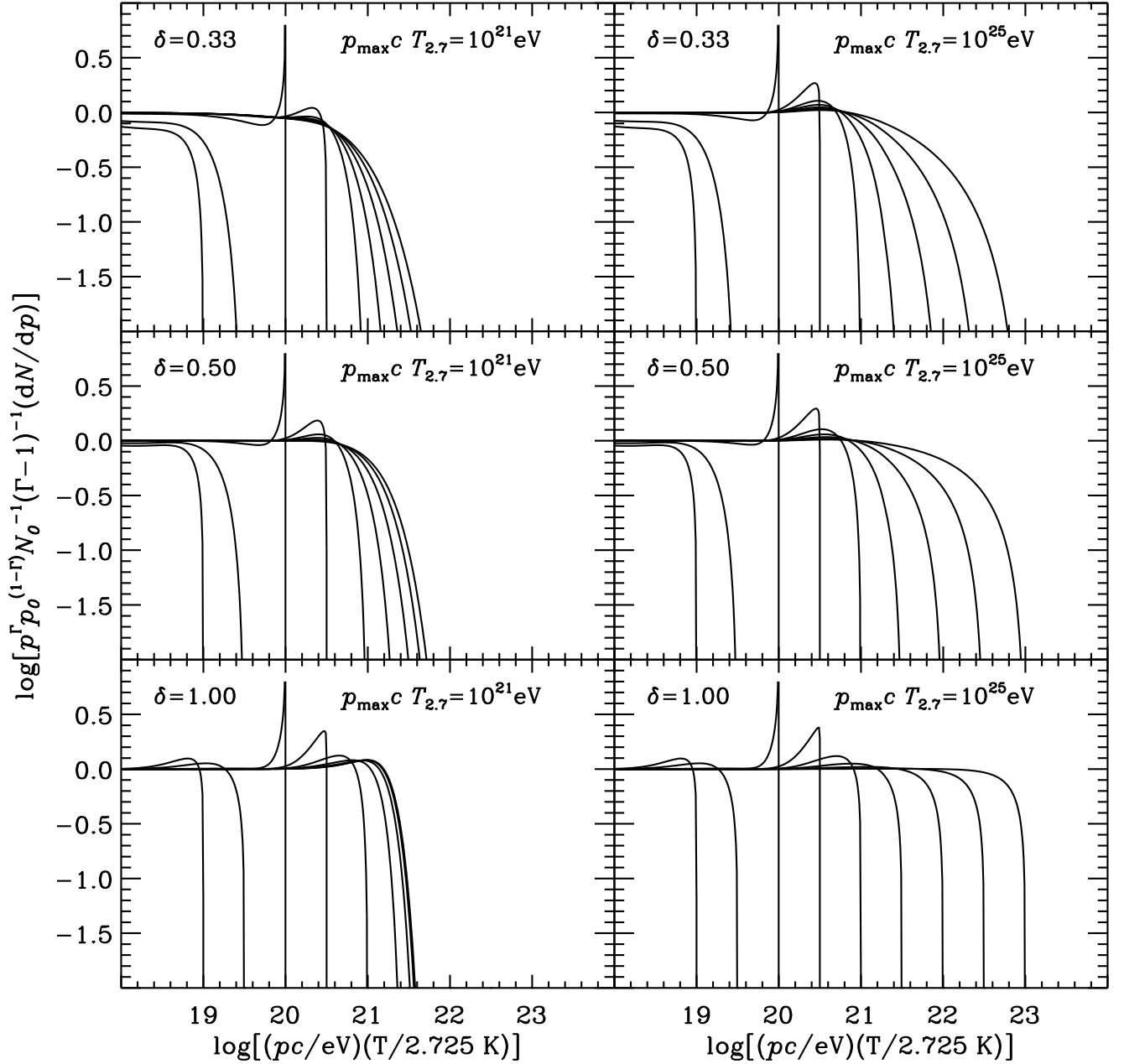,width=250mm}}
\caption{Differential momentum spectra of protons accelerated in
the CMBR with continuous losses by redshifting, Bethe-Heitler
pair production and pion photoproduction for $\Gamma=2$, and for
$\delta$ and $p_{\rm max}c$ as indicated.  In each case results
are shown for $p_{\rm cut}c= 10^{19}$ (leftmost curve),
$10^{19.5}$, \dots , $10^{23}$~eV (rightmost curve).}
\label{photoprod_cutoff_analytic}
\end{figure}

Treating pion photoproduction as a continuous loss process, the
escape rate and acceleration rate are given by
\begin{eqnarray}
r_{\rm esc}(p)  &=& (\Gamma-1)r_{\rm loss}(p_{\rm cut})\left[\left({p \over p_{\rm cut}}\right)^{-\delta} +
 \left({p_{\rm max} \over p_{\rm cut}}\right)^{-\delta}\right]+\delta \ell_2 r_{\rm loss}(p) \\
r_{\rm acc}(p) &=&  r_{\rm loss}(p_{\rm cut})\left({p \over p_{\rm cut}}\right)^{-\delta} - r_{\rm loss}(p)
\end{eqnarray}
and
the spectral index is 
\begin{eqnarray}
{d \ln F_{\rm esc} \over d \ln p}  &=& {\delta \ell_2 R(p)q(p) - \delta \, (\Gamma-1) x^{-\delta}  \over \delta \ell_2 R(p) + (\Gamma-1) (x^{-\delta} \, +  \, x_{\rm max}^{-\delta})} +{ \delta  x^{-\delta} + R(p)q(p)  \over x^{-\delta} - R(p)} \nonumber \\ && \,  - \,  1 \,  - \,  {\delta \ell_2 R(p) + C (x^{-\delta} + x_{\rm max}^{-\delta}) \over x^{-\delta} - R(p)}
\label{eq:slope_photo}
\end{eqnarray}
where   
\begin{eqnarray}
R(p) = {r_{\rm loss}(p) \over r_{\rm loss}(p_{\rm cut})}, \;\; 
q(p)={d \; \ln(r_{\rm loss}) \over  d \; \ln(p)}.
\end{eqnarray}
By examining the gradient of the solid curve in
Fig.~\ref{photoprod_rate}(a) one sees that $q(p)$ has a strong
peak at $\sim 10^{20}$~eV, and this will give rise to a strong
pile-up in the spectrum if the cut-off momentum is close to $\sim
10^{20}$~eV/c.  The resulting spectra for the continuous loss approximation are shown in
Fig.~\ref{photoprod_cutoff_analytic}, and do indeed show this
strong pile-up if $p_{\rm cut} \sim 10^{20}$~eV/c.  Pile-ups also
occur near other $p_{\rm cut}$ values, particularly if $p_{\rm
max} \gg p_{\rm cut}$, and are strongest if $\delta=1$.  However,
when the constant escape term (corresponding to, e.g., escape due
to finite size of accelerator) dominates escape near $p_{\rm
cut}$, then the resulting steepening in the spectrum is the
prominent feature in all cases (except where $p_{\rm cut} \sim
10^{20}$~eV/c when there is always a strong pile-up).  Of course,
the above treatment assumes Bethe-Heitler pair production and
pion photoproduction to be continuous loss processes.  While this
is a reasonable approximation for Bethe-Heitler pair production,
it certainly is not for pion photoproduction, and in this case
the sharp cut-off and pile-up will be partly smeared.  Monte
Carlo simulations have been performed to see to what extent any
pile-ups shown in Fig.~\ref{photoprod_cutoff_analytic} are washed
out.

In the Monte Carlo simulations, redshifting and Bethe-Heitler pair
production were treated as continuous loss processes, but
hadronic collisions of protons and neutrons were treated by the
Monte Carlo method using the SOPHIA event generator
\cite{Muecke00}.  Neutrons produced, e.g.\ in $p\gamma \to
n\pi^+$, may decay inside the box into protons which continue to
be accelerated, or they may escape from the box to decay outside
into cosmic rays depending on the neutron's Lorentz factor and
the physical dimensions of the acceleration region.  Since the
simulations are for a given $p_{\rm cut}$, $p_{\rm max}$ and
$\delta$, and the size of the box depends on the (unknown)
normalization of the diffusion coefficients, I have performed
simulations separately for the two extreme cases.  In the first
case the cosmic ray injection spectrum is made up only of protons
escaping from the box, whereas in the second case the cosmic ray
injection spectrum is made up of all neutrons produced inside the
box plus protons escaping from the box.  High energy gamma rays
and neutrinos also result from photoproduction during
acceleration (see ref.\ \cite{Szabo94} for acceleration of
protons and production of neutrinos during acceleration in
$\sim$$10^5$~K blackbody radiation).

\begin{figure}[htb]
\centerline{\hspace*{2cm}\epsfig{file=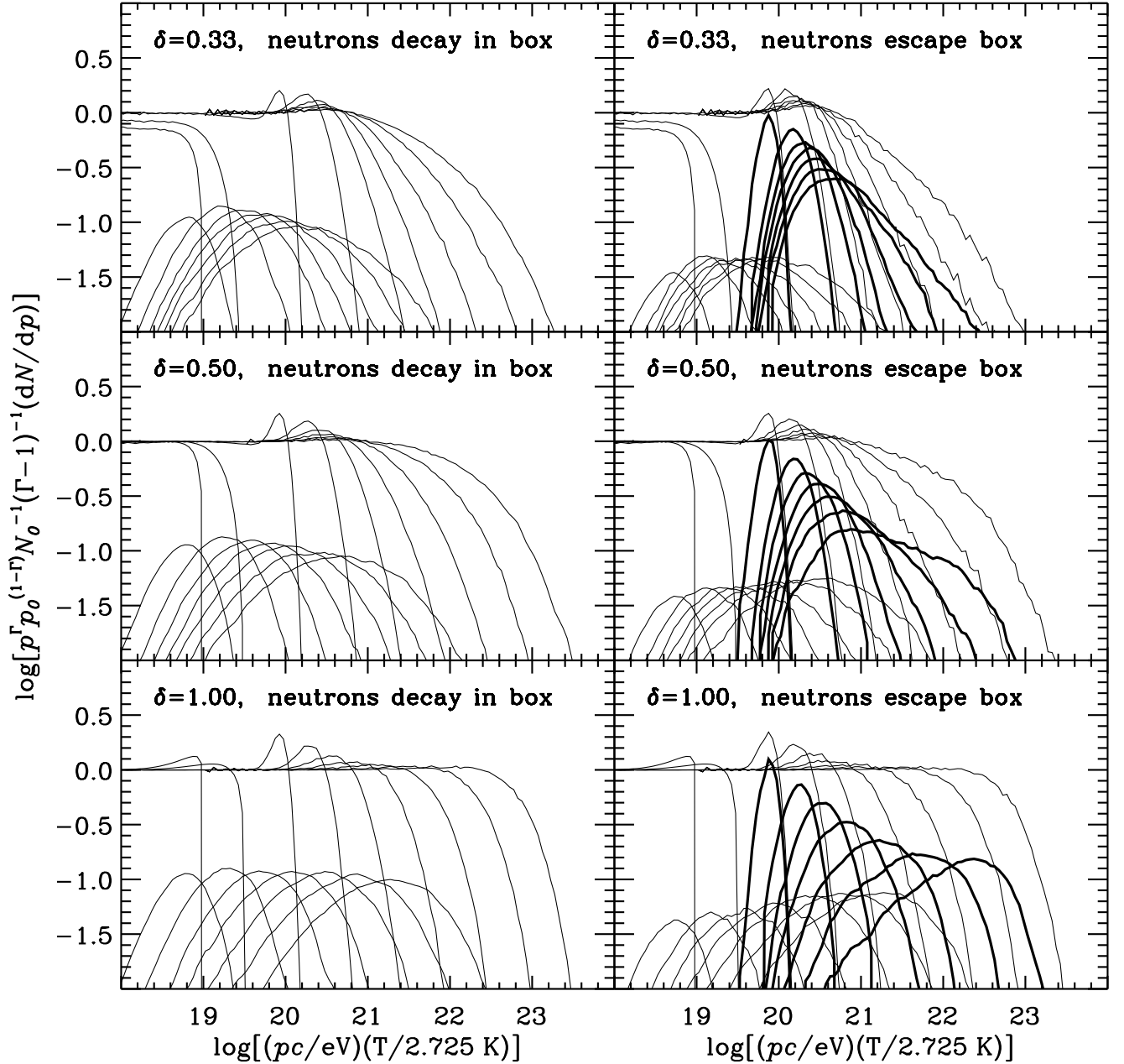,width=250mm}}
\caption{Differential momentum spectra of protons accelerated in
the CMBR with continuous losses by Bethe-Heitler pair production
and redshifting, and pion photoproduction interactions simulated
using the SOPHIA event generator for $\Gamma=2$, $p_{\rm max}c=
10^{25}$~eV, $\delta$ as indicated, and for neutron decay inside
the box (left column) and neutron escape from the box (right
column).  Upper thin curves show the proton flux for $p_{\rm
cut}c= 10^{19}$ (leftmost curve), $10^{19.5}$, \dots ,
$10^{23}$~eV (rightmost curve).  Lower thin curves total neutrino
flux (all flavours), and thick curves show spectra of escaping
neutrons, in both cases for $p_{\rm cut}c= 10^{20}$ (leftmost
curve), $10^{20.5}$, \dots , $10^{23}$~eV (rightmost curve). }
\label{photoprod_cutoff_mc}
\end{figure}

The Monte Carlo results for proton acceleration in a black body
radiation field with $\Gamma=2$ and $T_{\rm 2.7}p_{\rm max}c=
10^{25}$eV are shown in Fig.~\ref{photoprod_cutoff_mc}, with the
left column being for neutrons decaying inside the box, and the
right column being for neutrons escaping from the box before
decaying.  Upper thin curves are for escaping protons, and lower
thin curves are for neutrinos (all flavours) produced during
acceleration.  Comparing the $T_{\rm 2.7}p_{\rm max}c=10^{25}$eV
case (right column in Fig.~\ref{photoprod_cutoff_analytic}) with
the case of neutron decay inside the box (left column in
Fig.~\ref{photoprod_cutoff_mc}) confirms the presence of the
general features expected from treating interactions as a
continuous loss process.  The sharp features of
Fig.~\ref{photoprod_cutoff_analytic} are somewhat rounded, and
the spectrum contains particles with momenta exceeding $p_{\rm
cut}$, but this is to be expected because of the distributions of
inelasticity and interaction time.  

If neutrons escape from the
box before decaying, the protons from neutron decay will not be
further accelerated and this explains why the spectra in this
case are lower than the case of neutrons decaying inside the box.
In some models, protons are trapped outside of the acceleration
region but inside the astrophysical source, and so it is of
interest to know separately the spectrum of escaping neutrons produced
during the acceleration process (there will also be  be a
component due to pion photoproduction of the trapped accelerated  protons).
The contribution of escaping neutrons to the total escaping
proton flux is show in Fig.~\ref{photoprod_cutoff_mc} by the
thick curves. 

For the first two spectra, with $p_{\rm cut}c=1$--$3\times
10^{19}$eV, the cut-off is due to Bethe-Heitler pair production,
and so there is no significant neutrino or neutron component.  If
$p_{\rm cut}c=1$--$3\times 10^{20}$eV, there will be a strong
pile-up.  For $p_{\rm cut}c > 3\times 10^{20}$eV, the nature of
the spectrum above $3\times 10^{20}$eV depends strongly on the
momentum dependence of the diffusion coefficient.  For $\kappa
\propto p^{1/3}$, and neutrons escaping, the spectrum essentially
breaks from $\sim p^{-2}$ below the pile-up feature, to $\sim
p^{-4}$--$p^{-3}$ above the pile-up feature (corresponding to
$p_{\rm cut}c = 10^{21}$--$10^{23}$eV).  If neutrons decay inside
the box, the flux above the pile-up feature is higher and more
curved.  Nevertheless, for $p_{\rm cut}c > 3\times 10^{20}$eV and
with $\kappa \propto p^{1/3}$ the flux at $p \sim p_{\rm cut}$ is
substantially lower than expected without interactions.  In other
words, the effect of interactions cuts in at much lower momenta
than $p_{\rm cut}$, and this is most pronounced for $\kappa
\propto p^{1/3}$.  For $\kappa \propto p$ the spectra above
$3\times 10^{20}$eV generally continue as roughly the original
$\sim p^{-2}$ power law and suffer an almost exponential cut-off
at $p \sim p_{\rm cut}$, with the cut-off occurring at a slightly
lower momentum in the case of neutron escape.  As would be
expected, the spectra for $\kappa \propto p^{1/2}$ are
intermediate between the the $p^{1/3}$ and $p^1$ cases.

\subsection{Spectrum after propagation}

\begin{figure}[htb]
\centerline{\hspace*{2cm}\epsfig{file=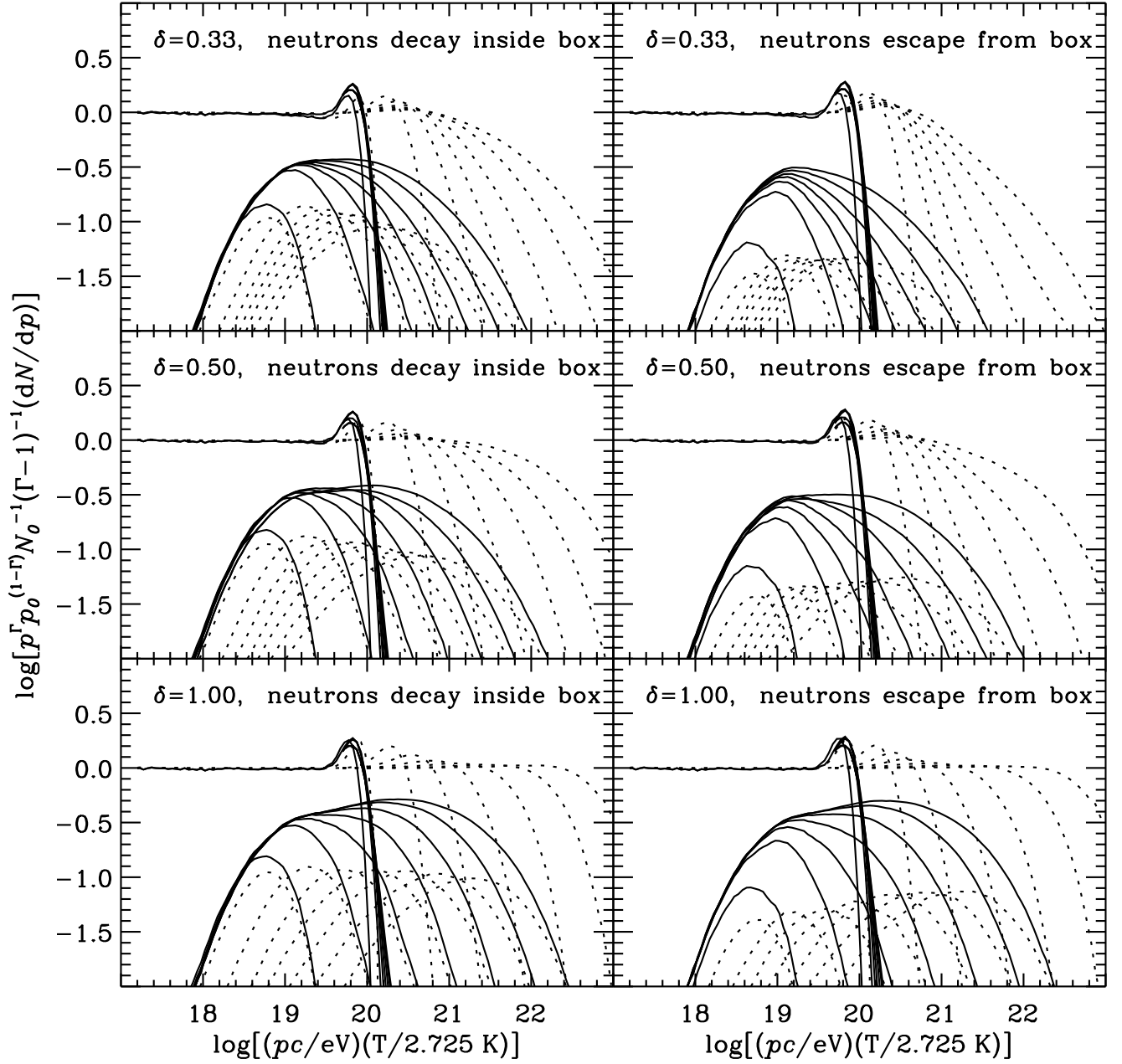,width=250mm}}
\caption{Spectra of protons and neutrinos (all flavours) escaping
from the acceleration region (dotted curves) and after
propagation for time 100~Mpc/c (solid curves) for $p_{\rm cut}c= 10^{20}$
(leftmost curves), $10^{20.5}$, \dots , $10^{23}$~eV (rightmost
curves) for the cases given in Fig.~\protect\ref{photoprod_cutoff_mc}.}
\label{evolv_photoprod_cutoff_mc}
\end{figure}

Unless the source is very near by, and the propagation time is
less than $\sim$10~Mpc/c, then the observed spectrum will be
severely affected by propagation.  This is illustrated in
Fig.~\ref{evolv_photoprod_cutoff_mc} which shows spectra of
protons and neutrinos (all flavours) escaping from the
acceleration region (dotted curves) and after propagation for
time 100~Mpc/c for the cases given in
Fig.~\ref{photoprod_cutoff_mc}.  Results (not shown) for 50~Mpc/c
are only marginally different, implying that cascading is
essentially complete by 100~Mpc/c.  The proton
spectra after propagation for all cases are hardly distinguishable, as they all
show, as expected, a prominent GZK cut-off at $10^{20}$eV with a
pile-up just below the acceleration cut-off.  Hence, it appears
that one can conclude little about the acceleration cut-off
energy and hence physical conditions of the acceleration region,
by directly observing the spectrum of UHE CR.  

\begin{figure}[htb]
\centerline{\epsfig{file=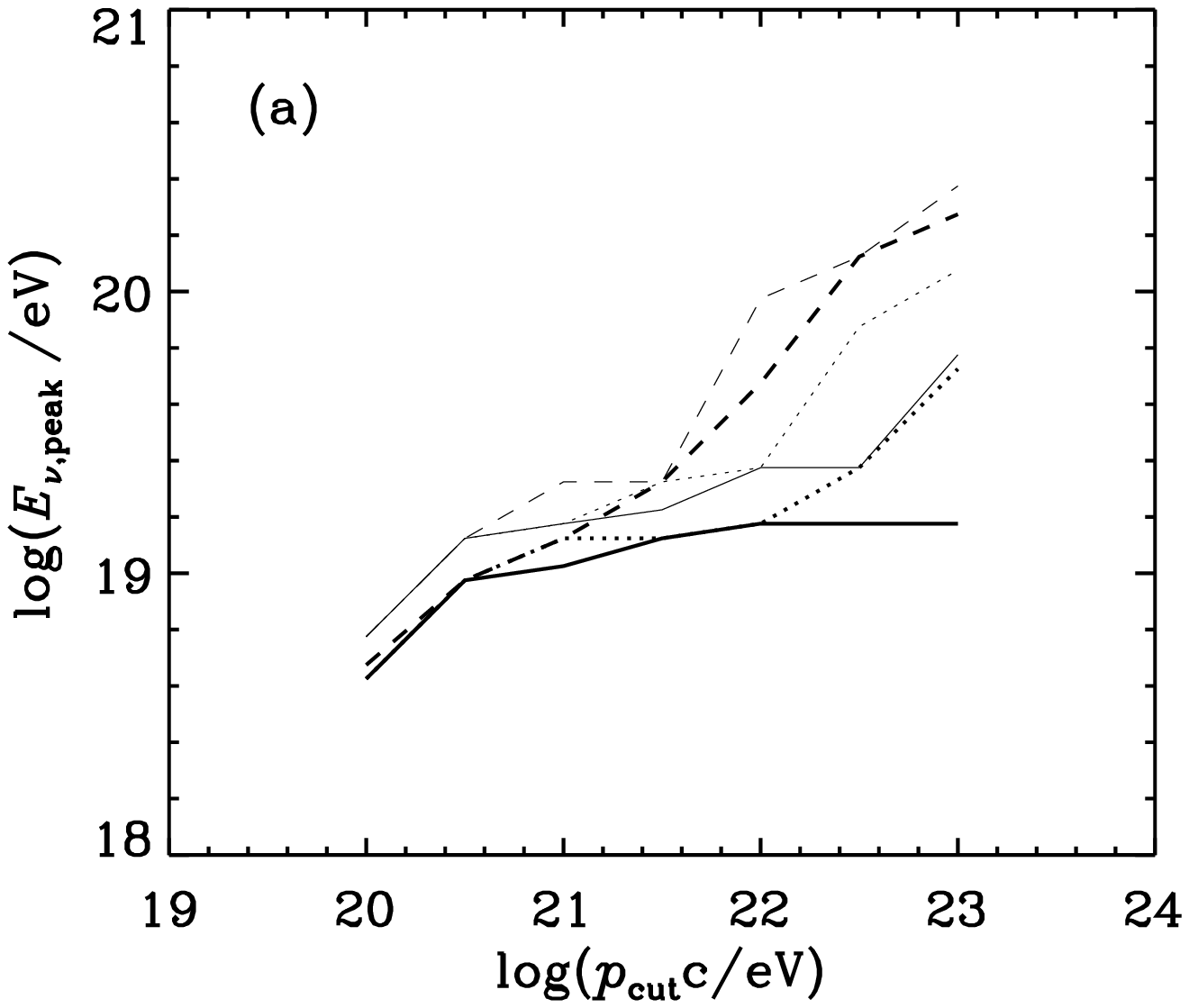,width=110mm}\hspace*{-4em}\epsfig{file=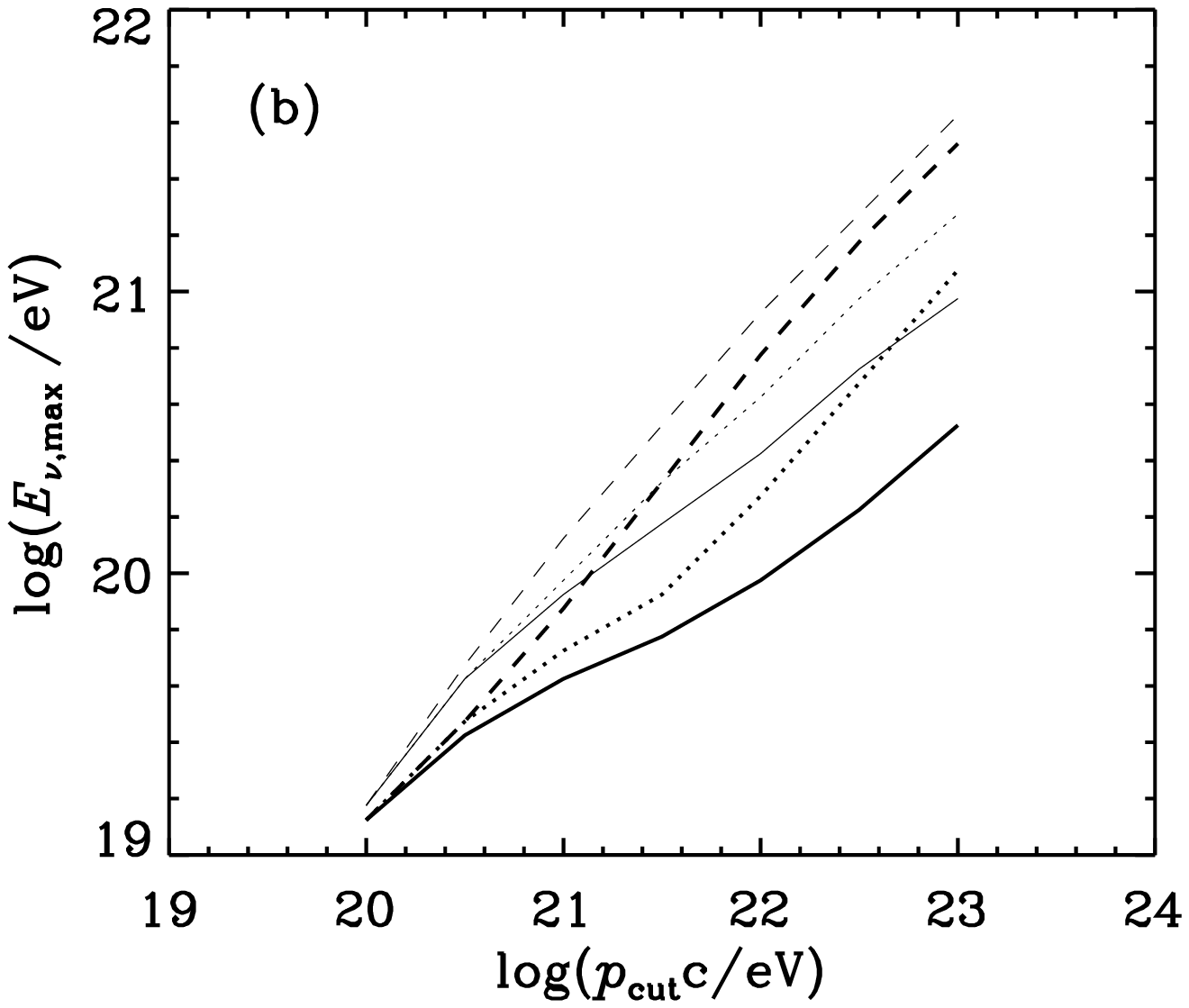,width=110mm}}
\caption{(a) Energy at peak of neutrino SED after propagation for
time 100~Mpc/c vs.\ proton cut-off momentum, and (b) maximum
neutrino energy defined as the energy at which the neutrino SED,
after propagation for time 100~Mpc/c, is $1/e$ of its peak value
vs.\ proton cut-off momentum.  Thin curves are for neutron decay
inside acceleration region, and thick curves are for neutron
escape from acceleration region; results are shown for
$\delta$=1/3 (solid curves), 1/2 (dotted curves) and 1 (dashed
curves).}
\label{evolv_numax_pcut}
\end{figure}

Information is preserved in the neutrino spectrum, to some
extent, about the proton cut-off energy and the momentum
dependence of the diffusion coefficient during acceleration.
During propagation, the energy of protons above the GZK cut-off
is converted into secondary particles, including neutrinos which
propagate essentially unhindered.  As can be seen, the cut-off in
the neutrino spectrum after propagation is directly related to
the cut-off in the accelerated proton spectrum, and the spectral
shape retains some information about the shape of the cut-off of
the accelerated proton spectrum, and hence the physical
conditions of the accelerator.  The spectra plotted are
effectively $E^2dN/dE$ and so are proportional to the energy flux
of particles per logarithmic energy interval, or the ``spectral
energy distribution'' (SED).  In Fig.~\ref{evolv_numax_pcut}(a) I
plot the energy at the peak of neutrino SED after propagation
vs.\ proton acceleration cut-off momentum, and in
Fig.~\ref{evolv_numax_pcut}(b) the maximum neutrino energy, which
I define here as the energy at which the neutrino SED is $1/e$ of
its peak value, vs.\ proton acceleration cut-off momentum.  As
can be seen, the maximum neutrino energy is a more useful
measurement for determining the proton acceleration cut-off
momentum as it is less subject to statistical fluctuations (in
real or Monte Calo data) and is more sensitive to $p_{\rm cut}$.

The inferred proton acceleration cut-off momentum is sensitive to
whether or not the neutrons escape from the acceleration region,
particularly for the case of a Kolmogorov spectrum of turbulence.
However, one can still use
Fig.~\ref{evolv_numax_pcut}(b) to put limits on the proton
cut-off energy.  For example, the observation of UHE neutrinos
with $E_{\nu, \rm max}=3\times 10^{20}$eV would imply a proton
cut-off energy of 3--4$\times 10^{21}$eV ($\delta$=1), 5$\times
10^{21}$--$10^{22}$eV ($\delta$=1/2), and 1.5$\times
10^{22}$--$10^{23}$eV ($\delta$=1/3).

\section{Conclusion}

In this paper I have discussed the shape of the high-energy part
of the spectrum of particles accelerated by DSA in the presence
of (i) additional diffusive escape from the accelerator, (ii)
continuous energy losses, (iii) energy changes arising from
interactions.  I find that the shapes of spectral features
(roll-offs, pile-ups, cut-offs) depend strongly on the
momentum-dependence of the acceleration rate, which in turn
depends on the momentum-dependence of the diffusion coefficient,
$\kappa \propto p^\delta$, and on the physical process
responsible for inhibiting acceleration.  Hence, the shape of the
high-energy part of the spectrum is important because it can give
clues to physical conditions at the acceleration site.  In some
situations the accelerated spectrum could, in principle, be
observed directly, but more usually the particle spectrum (if
observed) would be affected by propagation from source to
observer.  Nevertheless, some features of the spectrum after
acceleration could remain, or the spectrum of radiation emitted
by the accelerated particles would retain some sensitivity to the
features of the accelerated particle spectrum.

In the case of continuous losses where the spectrum is cut off at
the momentum at which acceleration rate equals loss rate,
pile-ups occur for $p^2$-losses for $\Gamma \le 2$, and for
$\Gamma$=2.5 if $\delta=1$.  However, for $p^1$-losses, there is
a fairly strong smooth cut-off except for the flattest spectrum
considered, $\Gamma$=1.5 with $\delta=1$. In the case of
additional escape processes inhibiting acceleration at momenta
comparable to or below the cut-off momentum, no pile-ups are seen,
and in all cases the spectra roll off smoothly at the momentum
where the additional escape becomes comparable with the normal
escape downstream.  One application of this could be in detailed
fitting the X-ray spectra of SNR \cite{DoneaProtheroe2004}, or other sources
in which synchrotron radiation is observed.  Another example
where these subtleties could be important is in acceleration in
AGN jets where, for some physical conditions, the spectrum could
be cut-off either by adiabatic losses as the jet expands, or by
leakage out of the jet.  In the jet-frame a reasonable order of
magnitude approximation for the time scales of both processes is
the jet radius divided by $c$.  Depending on which process
dominates, and the physical parameters, the spectrum could be
strongly cut off with spectrum affected well below the nominal
cut-off momentum (adiabatic losses), or have a more gentle
decline staring just before the nominal cut-off momentum and
extending a bit above it (diffusive escape).

In the case of interactions, the fate of the spectral features
depends largely on the mean inelasticity of the interaction.  For
very small inelasticity, the results are close to the
equivalent continuous loss case, but if 
$\bar{\alpha} \sim 1$ pile-ups and cut-offs are smoothed and particle
spectra extend beyond the nominal cut-off.  
I have considered in
detail the case of pion photoproduction by protons on black
body radiation, focusing on acceleration of UHE CR in the
presence of the CMBR.  Using plausible acceleration
rate parameters, I have re-examined the physical conditions
necessary to accelerate protons to UHE and, by considering a
``Hillas plot'' with constraints from photoproduction and
synchrotron radiation added, confirm that the most likely of the
originally proposed acceleration sites are in hot-spots in lobes
of giant radio galaxies, as inferred by Hillas \cite{Hillas84} and others
subsequently.  However, relativistically moving sources such as
AGN jets and GRB do not need to accelerate particles directly to
UHE as Doppler boosting from the plasma frame, in which the
acceleration occurs, to the host-galaxy frame can increase the
energies of escaping particles, for example, by $\sim$30 for AGN
jets and $\sim$1000 for GRB.  Thus, even though these sources
appear to be ruled out on the modified Hillas plot, they could
well be viable.

If the UHE CR are indeed accelerated, and do not result from a
top-down scenario, then the high energy end of the accelerated
particle spectrum could contain clues of the physical conditions
of the acceleration region.  The spectra on acceleration for
different physical conditions are quite distinct.  If the source
of the UHE CR is Galactic, e.g.\ a GRB which occurred in our
Galaxy in recent pre-historic times (say within a few tens of
thousands of years), then one might directly observe pile-up
features, possibly at super-GZK energies.  If the sources are
extragalactic the particle spectrum one would observe would
almost certainly have had all information about the part of the
spectrum above $\sim10^{20}$eV, i.e.\ beyond the GZK cut-off,
removed.  However, UHE neutrinos produced during propagation can
be used to estimate at least the cut-off energy of the protons
just after acceleration, if some reasonable assumptions about
momentum dependence of diffusion coefficients and size of the
accelerator can be made.  The neutrinos produced during
propagation will not point back directly to the acceleration
region, but will appear to come from a halo of size somewhere
between the proton gyroradius in intergalactic space around the
source and the pion photoproduction loss distance,i.e.\ $\sim$1
to 15~Mpc, depending on the magnetic field environment around the
source.  At the 16~Mpc distance of M87 this would mean neutrinos
would arrive from this source at angles up to
$4^\circ$--$90^\circ$, whereas for Cen~A at only $\sim$4 Mpc, one
might directly observe cut-off features in the UHE CR spectrum,
but one would observe only those neutrinos from the direction of
Cen~A produced during acceleration unless the accelerated cosmic
rays are efficiently trapped near to Cen~A.  For more distant
source candidates, it might be possible to identify some
currently active sources of UHE CR. Finally, I mention that if
heavy nuclei are accelerated then photodisintegration may take
place during acceleration and as well as during subsequent
propagation.  Simulations are planned for this case and will be
the subject of a future paper.

\section*{Acknowledgments}

I thank Alina Donea and Greg Thornton for carefully reading the
manuscript.  This research is supported by a Discovery Project
grant awarded by the Australian Research Council.



\newpage


\newpage

\begin{thebibliography}{90}

\bibitem{Drury83a} L.O'C. Drury, Space Sci. Rev. 36 (1983) 57

\bibitem{BlandfordEichler87} R. Blandford, D. Eichler, Phys. Rep. 154 (1987) 1

\bibitem{BerezhkoKrymsky88} E.G. Berezhko,  G.F. Krymski, Usp. Fiz. Nauk 154 (1988) 49

\bibitem{JonesEllison91} F.C. Jones, D.C. Ellison,  Space Sci. Rev. 58 (1991) 259

\bibitem{ProtheroeStanev98} R.J. Protheroe and T. Stanev, Astropart. Phys. 10 (1999) 185

\bibitem{greisen} K. Greisen,  Phys. Rev. Lett. 16 (1966) 748

\bibitem{zatsepin} G.T. Zatsepin, V.A Kuzmin, Sov. Phys. JETP Lett. 4 (1966) 78

\bibitem{rf:TAK} M. Takeda et al., Astropart. Phys. 19 (2003) 447

\bibitem{rf:FE3} D.J. Bird et al.,  ApJ 441 (1995) 144

\bibitem{HiRes02} T. Abu-Zayyad et al., Astropart. Phys., submitted astro-ph/0208301

\bibitem{ProtheroeJohnson95} R.J. Protheroe and P.A. Johnson, Astroparticle Physics, 4, 253 (1996); erratum,  5, 215 (1996)

\bibitem{Stanevetal00}T. Stanev, R. Engel, A. M\"ucke, R.J. Protheroe, 
	J.P. Rachen, Phys. Rev. D 62 (2000) 093005 

\bibitem{Mannheimetal01} K. Mannheim, R.J. Protheroe, and J.P. Rachen,
	Phys. Rev. D 63  (2001) 023003

\bibitem{Hayashidaetal99}  N. Hayashida et al., Astropart. Phys. 10 (1999) 303

\bibitem{AugerCollaboration2001} Auger Collaboration Contributions 2001, in Proceedings of XXVII Cosmic Ray Conference, Hamburg, Edited by M. Simon et al., Copernicus Gesellschaft 2001, Vol. 2, pp 699-787

\bibitem{NaganoWatson00} M. Nagano, A.A. Watson, Rev. Mod. Phys. 72  (2000) 689 

\bibitem{ProtheroeClay04}R.J. Protheroe and R.W. Clay,
	Pub. Astron. Soc. Austr. (2004), in press. astro-ph/0311466 

\bibitem{CasseMarkowith03} F. Casse, A. Markowith, Astron. Astrophys. 404 (2003) 405

\bibitem{Druryetal99} L.O'C. Drury, P. Duffy, D. Eichler, A. Mastichiadis, Astron. Astrophys. 347 (1999) 370

\bibitem{MueckeProtheroe01} A. M\"ucke, R.J. Protheroe, Astroparticle Physics 15 (2001) 121

\bibitem{Mueckeetal03} A. M\"ucke, R.J. Protheroe, R. Engel, J.P. Rachen,
	T. Stanev, Astropart. Phys. 18 (2003) 593

\bibitem{Hillas84}  A.M. Hillas, Ann.~Rev.~Astron.~Astrophys., 22 (1984) 425

\bibitem{RachenBiermann93} J.P. Rachen, P.L. Biermann, Astron. Astrophys. 272 (1993) 161

\bibitem{Schopperetal02} R. Schopper, G. Thorsten Birk, 
 H. Lesch,  Astropart. Phys. 17 (2002) 347

\bibitem{Normanetal95} C.A. Norman,  D.B. Melrose, A. Achterberg, ApJ 454 (1995) 60

\bibitem{Protheroe00}  R.J. Protheroe, in ``Topics in cosmic-ray astrophysics'' ed. M.A.  DuVernois, Nova Science Publishing: New York, 2000, pp 258--298

\bibitem{BiermannStrittmatter87} P.L. Biermann,, \&  P.A.  Strittmatter, ApJ 322 (1987)  643 

\bibitem{Medvedev03}  M.V.Medvedev, Phy. Rev. E 67 (2003) 045401

\bibitem{Protheroeetal03} R.J. Protheroe, A.-C. Donea, A. Reimer, Astropart. Phys. 19 (2003) 559

\bibitem{PelletierKersate00} G. Pelletier, E. Kersal\'e, Astron. Astrophys. 361 (2000) 788

\bibitem{Haswelletal92} C.A. Haswell,  T. Tajima,  J.-L. Sakai,  ApJ 401 (1992) 495

\bibitem{Sorrell87} W.H. Sorrell, ApJ 323 (1987) 647

\bibitem{LitwinRosner01} C. Litwin, R. Rosner, Phys. Rev. Lett. 86 (2001) 4745

\bibitem{deGouvelaDalPinoLazarian01} E.M. de Gouvela Dal Pino, A. Lazarian, ApJ 560 (2001) 358

\bibitem{Arons03} J. Arons, Astrophysical Journal 589 (2003) 871

\bibitem{Muecke00} A. M\"ucke, R. Engel, J.P. Rachen, R.J. Protheroe, T. Stanev, 
	Comp. Phys. Com. 124 (2000) 290

\bibitem{Szabo94} A.P. Szabo, R.J. Protheroe, Astropart. Phys. 2 (1994) 375

\bibitem{DoneaProtheroe2004} A.-C. Donea, R.J. Protheroe, in preparation (2004)


\end{thebibliography}
\end{document}